\documentclass[aps,showpacs,preprintnumbers,twocolumn,superscriptaddress,floatfix,nofootinbib]{revtex4-2}

\usepackage[linesnumbered,lined,boxed,commentsnumbered,ruled,vlined]{algorithm2e}
\usepackage{algpseudocode}
\usepackage{amsfonts}
\usepackage{amsmath}
\usepackage{amssymb}
\usepackage{amsthm}
\usepackage{bm,bbm}
\usepackage{caption}
\usepackage{comment}
\usepackage{dcolumn}
\usepackage{epstopdf}
\usepackage{float}
\usepackage{graphicx}
\usepackage{hyperref}
\usepackage{cleveref}
\usepackage{mathrsfs}
\usepackage{mathtools}
\usepackage{multirow}
\usepackage{physics}
\usepackage{ragged2e}
\usepackage{scalerel}
\usepackage{subfigure}
\usepackage{tabularx}
\usepackage{tikz}
\usepackage{url}
\usepackage{utfsym}
\usepackage{verbatim}
\usepackage{xcolor}
\usepackage{xkcdcolors}
\usepackage[normalem]{ulem}
\usepackage{lineno}
\DeclareMathOperator*{\argmax}{arg\,max}

\DeclareCaptionJustification{justified}{\justifying}
\hypersetup{colorlinks=true, citecolor=orange, urlcolor=blue, linkcolor=magenta}

\definecolor{cel}{rgb}{0.0,0.53,0.74}
\definecolor{green}{rgb}{0.0,0.5,0.0}

\definecolor{colSDRG}{RGB}{51,117,56}
\definecolor{colSDCM}{RGB}{194,106,119}
\definecolor{colRec}{RGB}{230,158,0}
\definecolor{colpp}{RGB}{87,181,232}
\definecolor{colnn}{RGB}{0,158,115}
\definecolor{colpn}{RGB}{204,120,166}

\begin{document}

\title{A Bayesian approach to out-of-sample network reconstruction}

\author{Mattia Marzi}
\email{mattia.marzi@imtlucca.it}
\affiliation{IMT School for Advanced Studies, P.zza San Francesco 19, 55100 Lucca (Italy)}
\affiliation{INdAM-GNAMPA Istituto Nazionale di Alta Matematica `Francesco Severi', P.le Aldo Moro 5, 00185 Rome (Italy)}
\author{Tiziano Squartini}
\affiliation{IMT School for Advanced Studies, P.zza San Francesco 19, 55100 Lucca (Italy)}
\affiliation{Scuola Normale Superiore, P.zza dei Cavalieri 7, 56126 Pisa (Italy)}
\affiliation{INdAM-GNAMPA Istituto Nazionale di Alta Matematica `Francesco Severi', P.le Aldo Moro 5, 00185 Rome (Italy)}

\date{\today}

\begin{abstract}
Networks underpin systems that range from finance to biology, yet their structure is often only partially observed. Current reconstruction methods typically fit the parameters of a model anew to each snapshot, thus offering no guidance to predict future configurations. Here, we develop a Bayesian approach that uses the information about past network snapshots to inform a prior and predict the subsequent ones, while quantifying uncertainty. Instantiated with a single-parameter fitness model, our method infers link probabilities from node strengths and carries information forward in time. When applied to the Electronic Market for Interbank Deposit across the years $1999$-$2012$, our method accurately recovers the number of connections per bank at subsequent times, outperforming probabilistic benchmarks designed for analogous, link prediction tasks. Notably, each predicted snapshot serves as a reliable prior for the next one, thus enabling self-sustained, out-of-sample reconstruction of evolving networks with a minimal amount of additional data.
\end{abstract}

\maketitle

\section*{INTRODUCTION}

Network theory is employed to address problems of scientific and societal relevance, from the prediction of epidemic spreading to the identification of early-warning signals of upcoming financial crises~\cite{Colizza2006,Barrat2008,Newman2010,Pastor2015,squartini2013early,battiston2016complexity,bardoscia2017pathways,macchiati2025spectral}. 

Any dynamical process is strongly affected by the topology of the underlying network; still, data restrictions may prevent it from being fully accessible - in many empirical settings, only aggregate node-level quantities are available. The inference problem we address here consists in turning the available information into a probability distribution over the unknown network portion, so as to quantify the plausibility of each link.

Such a problem can be approached by constructing a statistical benchmark, i.e. a set of configurations that preserve only certain properties while maintaining controlled randomness elsewhere: these properties, named \emph{constraints} and indicated as $\mathbf{C}$, represent the only information that can be employed to infer any other (inaccessible) property.

A class of models whose popularity has steadily increased over the years is that of Exponential Random Graphs (ERGs)~\cite{park2004statistical,Bianconi2007,squartini2011analytical,Fronczak2012,squartini2015unbiased,saracco2015randomizing,cimini2019statistical}. ERGs come from constrained Shannon entropy maximization~\cite{jaynes1957information,park2004statistical,squartini2011analytical,squartini2017maximum} and belong to the category of approaches enforcing constraints (only) on average.

So far, the parameters defining ERGs have been determined by employing the traditional maximum-of-the-likelihood estimation procedure~\cite{Handcock2003InferenceDegeneracy,garlaschelli2008maximum,VanDuijn2009MPLEvsMLE,vallarano2021fast,divece2023deterministic}; in the case of ERGs, in fact, it leads to the particularly simple request that the expected value of each constraint equals the corresponding empirical value: in symbols, $\langle\mathbf{C}\rangle=\mathbf{C}^*$, with $\langle\mathbf{C}\rangle$ indicating the expected value of the vector of constraints and $\mathbf{C}^*$ the empirical one.

Coming to the purpose of reconstructing a graph topology, one of the most effective models is the Undirected Binary Configuration Model (UBCM)~\cite{park2004statistical,squartini2011analytical,Fronczak2012,squartini2015unbiased,cimini2019statistical}, induced by the degree sequence $\{k_i\}_{i=1}^N$, with $k_i=\sum_{j(\neq i)}a_{ij}$ indicating the number of neighbors of node $i$, $a_{ij}$ indicating the entry $(i,j)$ of the adjacency matrix $\mathbf{A}$ of a binary undirected network and $N$ indicating the total number of nodes. More formally, the UBCM assigns $\mathbf{A}$ the factorized probability distribution reading

\begin{equation}\label{eq:P}
P(\mathbf{A}|\mathbf{x})=\prod_{i=1}^N\prod_{j(>i)}p_{ij}^{a_{ij}}(1-p_{ij})^{1-a_{ij}},
\end{equation}
where

\begin{equation}\label{eq:pij}
p_{ij}=\frac{x_ix_j}{1+x_ix_j}
\end{equation}
and $x_i$ indicates the Lagrange multiplier associated to $k_i$; maximizing the log-likelihood of such a model implies solving the system of equations

\begin{equation}
\langle k_i\rangle=\sum_{j(\neq i)}p_{ij}=k_i^*,\:\forall\:i
\end{equation}
that, in turn, induces the vector of point-wise estimates hereby indicated with $\{x_i^*\}_{i=1}^N$.

A similar perspective can be adopted to approach \emph{link prediction}, an active research field within the broader one of network science. Close in scope to network reconstruction~\cite{ReconstructionMethods2018}, link prediction targets specific connections, aiming to uncover missing ones and predict those most likely to emerge in the future~\cite{Zhou_2021,lu_link_2011,santucci_missing_2026}. Such an issue is relevant in many research areas, such as those concerning socio-economic and financial networks: knowing the structure of the commercial partnerships between firms or of the financial exchanges between banks is, in fact, relevant for a number of reasons, such as quantifying the risk associated to the propagation of a shock~\cite{bardoscia_physics_2021,ReconstructingFirmLevelInteractionsDutchInputOutputNetwork2022,mungo_reconstructing_2023}. Link prediction algorithms rank unconnected node pairs on the basis of a \emph{score}: while some methods rely on purely structural information, others admit external information as well.

The simplest framework to carry out link prediction includes the so-called \emph{similarity-based} algorithms: scores, here, are induced by some measure of similarity between nodes; to this aim, \emph{local}, \emph{quasi-local} or \emph{global} information - such as the degree, the degree of common neighbours or the length of paths connecting any two nodes - has been employed~\cite{Zhou_2021,lu_link_2011,santucci_missing_2026}.

A more refined framework includes machine learning approaches~\cite{Zhou_2021,lu_link_2011,santucci_missing_2026}: specifying a model, here, amounts to learning a function that determines the probability for nodes $i$ and $j$ to be connected while taking as inputs two vectors $\mathbf{f}_i$ and $\mathbf{f}_j$ of (structural and/or external) node-specific features and a vector $\mathbf{g}_{ij}$ of (structural and/or external) edge-specific features.

A third alternative is represented by the so-called \emph{likelihood-based} algorithms, defined by a likelihood function whose maximization provides the probability that any two nodes are connected, to be interpreted as a score for the link existence: this is achieved precisely by assuming that a certain amount of information is accessible, hence treating it as a constraint to account for~\cite{Zhou_2021,lu_link_2011,santucci_missing_2026}.

The third perspective justifies the interpretation of the link prediction problem as an instance of the network reconstruction one, although network reconstruction usually deals with less information to predict more aggregate properties; in this sense, the key distinction between these two approaches is informational: while link prediction exploits dyadic features to achieve an edge-level classification that is as accurate as possible, entropy-based reconstruction makes the best possible use of the available information to reproduce the largest possible portion of a network.\\

Let us explicitly notice that the schemes sketched above have been designed to carry out inference in an \emph{in-sample} fashion: in other words, the probability coefficients $\{p_{ij}\}_{i,j=1}^N$ are estimated from the available information concerning $\mathbf{A}^*$ to make statements about $\mathbf{A}^*$ itself. To the best of our knowledge, a genuinely \emph{out-of-sample} formulation of the entropy-based framework for network reconstruction is still missing: existing methods for temporal link prediction, in fact, address such a task under highly controlled circumstances such as being in presence of the same set of nodes, having full access to all sets of dyadic features, as well as to a stable representation of the latter, etc.

Here, instead, the aforementioned issue is addressed by developing a Bayesian framework in which past snapshots inform an empirical prior only through the sufficient statistics needed to calibrate the model - i.e. without requiring additional regularities that are not supported by data. When instantiated with a single-parameter model, our framework \emph{i)} allows the number of nodes to vary over time; \emph{ii)} allows one to reconstruct future snapshots from node-level information and propagated uncertainty (let us, in fact, notice that the derivation of ERGs offers no prescription to propagate the uncertainty affecting a parameter estimation from one snapshot to another, nor to turn past observations into an informative prior about future ones); \emph{iii)} in its self-sustained version, evolves without requiring any other topological information than the one needed for initialization; \emph{iv)} remains competitive when compared with an in-sample reconstruction model calibrated on the target snapshot, thus showing that an accurate out-of-sample reconstruction can be achieved with a limited amount of accessible, interpretable information.

\section*{RESULTS}

\subsection*{Towards a Bayesian approach}

To carry out an out-of-sample network reconstruction, let us, then, move towards a different design, trying to turn the UBCM into a Bayesian model. Bayes' theorem provides the key relation $P(\mathbf{x}|\mathbf{A})P(\mathbf{A})=P(\mathbf{A},\mathbf{x})=P(\mathbf{A}|\mathbf{x})\pi(\mathbf{x})$ which, in turn, induces the following definition of \emph{posterior distribution}:

\begin{equation}\label{eq:pd}
P(\mathbf{x}|\mathbf{A})=\frac{P(\mathbf{A}|\mathbf{x})\pi(\mathbf{x})}{P(\mathbf{A})}=\frac{P(\mathbf{A}|\mathbf{x})\pi(\mathbf{x})}{\int P(\mathbf{A}|\mathbf{x})\pi(\mathbf{x})d\mathbf{x}};
\end{equation}
adapting the parameter estimation procedure accordingly would amount to calculating either the \emph{mean} or the \emph{mode} of the posterior distribution, respectively defined as

\begin{equation}
\langle\mathbf{x}|\mathbf{A}\rangle=\int\mathbf{x}P(\mathbf{x}|\mathbf{A})d\mathbf{x}
\end{equation}
and

\begin{equation}
\hat{\mathbf{x}}_\text{MAP}=\argmax_{\mathbf{x}}\{P(\mathbf{x}|\mathbf{A})\},
\end{equation}
the acronym standing for \emph{maximum-a-posteriori}. Both procedures, however, provide a point-wise estimate solely depending on the observation made at time $t$. A better suited instrument seems to be the so-called \emph{posterior predictive distribution}~\cite{GelmanBDA3}. Upon calling $\mathbf{A}_t$ the adjacency matrix representing our graph at time $t$ and $\mathbf{A}_{t+1}$ the adjacency matrix representing our graph at time $t+1$, it reads

\begin{align}\label{eq:ppp}
P(\mathbf{A}_{t+1}|\mathbf{A}_t)&=\int P(\mathbf{A}_{t+1},\mathbf{x}|\mathbf{A}_t)d\mathbf{x}\nonumber\\
&=\int P(\mathbf{A}_{t+1}|\mathbf{x},\mathbf{A}_t)P(\mathbf{x}|\mathbf{A}_t)d\mathbf{x}\nonumber\\
&=\int P(\mathbf{A}_{t+1}|\mathbf{x})P(\mathbf{x}|\mathbf{A}_t)d\mathbf{x}\nonumber\\
&=\int\frac{P(\mathbf{A}_{t+1}|\mathbf{x})P(\mathbf{A}_t|\mathbf{x})\pi(\mathbf{x})}{P(\mathbf{A}_t)}d\mathbf{x},
\end{align}
where the chain rule $f(x,y|z)=f(x|y,z)f(y|z)$ has been employed at the second passage and conditional independence has been assumed at the third passage.

The multivariate nature of the formula above, however, makes its resolution cumbersome; moreover, the UBCM is not a viable reconstruction model, as its calibration requires the knowledge of the entire degree sequence per snapshot - a requirement that is hardly satisfied in a realistic scenario. For such a reason, let us look for a simpler, single-parameter model. Our choice, then, turns eq.~\ref{eq:ppp} into

\begin{align}
P(\mathbf{A}_{t+1}|\mathbf{A}_t)&=\int_0^{+\infty}\frac{P(\mathbf{A}_{t+1}|z)P(\mathbf{A}_t|z)\pi(z)}{P(\mathbf{A}_t)}dz,
\end{align}
where $P(\mathbf{A}_t)=\int_0^{+\infty}P(\mathbf{A}_t|z)\pi(z)dz$. Let us stress that assuming conditional independence simplifies the computation of the formula above, as the dependence of the future snapshot on the past snapshot is established only via $z$; moreover, nothing prevents us from considering a varying number of nodes, i.e. $N_t$ and $N_{t+1}$ can differ.

Since $z$ is integrated out, we are left with a probability depending on $\mathbf{A}_t$, that we know, and on $\mathbf{A}_{t+1}$, that we want to infer; the problem can be simplified even more upon considering that the marginal (i.e. edge-specific) probability reads

\begin{align}\label{eq:q}
q_{ij}^{t+1}&=P(a_{ij}^{t+1}=1|\mathbf{A}_t)=\int_0^{+\infty}p_{ij}^{t+1}(z)\frac{P(\mathbf{A}_t|z)\pi(z)}{P(\mathbf{A}_t)}dz,
\end{align}
where $p_{ij}^{t+1}$ is the probability that the corresponding entry of $\mathbf{A}_{t+1}$, i.e. $a_{ij}^{t+1}$, is $1$ (see also Appendix~\hyperlink{AppA}{A}).

Such an expression offers us a viable way to carry out our inference exercise, since it allows the following quantities to be evaluated analytically, i.e. the (expected) total number of links, reading

\begin{equation}
\langle L_{t+1}\rangle=\sum_{i=1}^N\sum_{j(>i)}q_{ij}^{t+1},
\end{equation}
the (expected) degree of each node, reading

\begin{equation}
\langle k_i^{t+1}\rangle=\sum_{j(\neq i)}q_{ij}^{t+1},\quad\forall\:i
\end{equation}
and the entries of the so-called \emph{confusion matrix}, i.e.

\begin{align}
\langle\text{TP}_{t+1}\rangle&=\sum_{i=1}^N\sum_{j(>i)}a_{ij}^{t+1}q_{ij}^{t+1},\\
\langle\text{FP}_{t+1}\rangle&=\sum_{i=1}^N\sum_{j(>i)}(1-a_{ij}^{t+1})q_{ij}^{t+1},\\
\langle\text{TN}_{t+1}\rangle&=\sum_{i=1}^N\sum_{j(>i)}(1-a_{ij}^{t+1})(1-q_{ij}^{t+1}),\\
\langle\text{FN}_{t+1}\rangle&=\sum_{i=1}^N\sum_{j(>i)}a_{ij}^{t+1}(1-q_{ij}^{t+1});
\end{align}
in words, $\langle\text{TP}_{t+1}\rangle$ represents the (expected) number of true positives, i.e. the number of correctly recovered connections; $\langle\text{FP}_{t+1}\rangle$ represents the (expected) number of false positives, i.e. the number of incorrectly recovered connections; $\langle\text{TN}_{t+1}\rangle$ represents the (expected) number of true negatives, i.e. the number of correctly recovered missing connections; $\langle\text{FN}_{t+1}\rangle$ represents the (expected) number of false negatives, i.e. the number of incorrectly recovered missing connections.

\subsection*{The Bayesian Erd\"os-R\'enyi Model (BERM)}

Let us, now, instantiate our framework with the Erd\"os-R\'enyi Model. Its classical formulation reads

\begin{equation}
P(\mathbf{A}|p)=p^{L(\mathbf{A})}(1-p)^{V-L(\mathbf{A})},
\end{equation}
where $V=N(N-1)/2$ indicates the total number of pairs of nodes. Fully determining it, however, requires the \emph{prior distribution} to be specified: a popular choice is that of considering the \emph{conjugate prior}, introduced to maintain the functional form of the likelihood function. Upon doing so, we are led to expression

\begin{align}
P(p|\mathbf{A})&=\frac{P(\mathbf{A}|p)\pi(p)}{\int_0^1P(\mathbf{A}|p)\pi(p)dp}\nonumber\\
&=\frac{p^{L(\mathbf{A})+\alpha-1}(1-p)^{V-L(\mathbf{A})+\beta-1}}{\text{B}(L(\mathbf{A})+\alpha,V-L(\mathbf{A})+\beta)},
\end{align}
coming from posing

\begin{align}
\pi(p)=\frac{p^{\alpha-1}(1-p)^{\beta-1}}{\text{B}(\alpha,\beta)},
\end{align}
where $\text{B}(\alpha,\beta)=\int_0^1p^{\alpha-1}(1-p)^{\beta-1}dp$ is the Beta function. As a consequence, we can write

\begin{align}
P(L_{t+1}=k|\mathbf{A}_t)&=\int_0^1P(L_{t+1}=k|p)P(p|\mathbf{A}_t)dp\nonumber\\
&=\int_0^1\binom{V_{t+1}}{k}p^k(1-p)^{V_{t+1}-k}\frac{P(\mathbf{A}_t|p)\pi(p)}{P(\mathbf{A}_t)}dp\nonumber\\
&=\text{BetaBin}(V_{t+1},L_t+\alpha,V_t-L_t+\beta),
\end{align}
where the explicit dependence of the third expression on $\mathbf{A}_t$ has been dropped; in words, the total number of links at time $t+1$, conditional to the observation of $\mathbf{A}_t$, obeys a beta-binomial distribution. Since

\begin{align}
\langle L_{t+1}|\mathbf{A}_t\rangle&=\sum_{k=0}^{V_{t+1}}kP(L_{t+1}=k|\mathbf{A}_t),
\end{align}
swapping the operations of sum and integration leads to

\begin{align}
\langle L_{t+1}|\mathbf{A}_t\rangle&=\int_0^1\sum_{k=0}^{V_{t+1}}kP(L_{t+1}=k|p)P(p|\mathbf{A}_t)dp\nonumber\\
&=V_{t+1}\int_0^1p\frac{P(\mathbf{A}_t|p)\pi(p)}{P(\mathbf{A}_t)}dp
\end{align}
and comparing it with eq.~\ref{eq:q} further leads us to recognize that

\begin{align}
q^{t+1}&=\int_0^1p\frac{P(\mathbf{A}_t|p)\pi(p)}{P(\mathbf{A}_t)}dp\nonumber\\
&=\frac{\text{B}(1+L_t+\alpha,V_t-L_t+\beta)}{\text{B}(L_t+\alpha,V_t-L_t+\beta)}\nonumber\\
\label{eq:q_berm}
&=\frac{L_t+\alpha}{V_t+\alpha+\beta}
\end{align}
(see also Appendix~\hyperlink{AppB}{B}).

\subsection*{The Bayesian Fitness Model (BFM)}

The main limitation of the BERM lies in its homogeneity: in other words, all nodes are treated equally. Since different nodes are known to play different roles, according to their structural importance, let us look for a single-parameter model, \emph{heterogeneous in nature.}

A very natural choice is that of considering the variant of the UBCM named \textit{density-corrected Gravity Model} (dcGM)~\cite{CimiModel2015,Mazzarisi2017LimitedInformation,Anand2018MissingLinks,Lebacher2019LostEdges,Ramadiah2020ReconstructingAndStressTesting,cimini2021reconstructing}, induced by node-specific fitnesses typically identified with the strengths $\{s_i\}_{i=1}^N$, where $s_i=\sum_{j(\neq i)}w_{ij}$ and $\mathbf{W}$ is the weighted adjacency matrix associated to our graph, while solely enforcing (a proxy of) the link density, defined as $c=2L/N(N-1)=2\sum_{i=1}^N\sum_{j(>i)}a_{ij}/N(N-1)$. For consistency, $\mathbf{A}=\Theta(\mathbf{W})$, where $\Theta(x)$ denotes the Heaviside step function, here defined as $\Theta(x)=1$ if $x>0$ and $\Theta(x)=0$ otherwise, applied element-wise. More quantitatively, the dcGM is defined by the fitness ansatz $x_i=\sqrt{z}s_i$, which exploits the empirical correlation between the Lagrange multiplier $x_i$ and the strength $s_i$, thus turning eq.~\ref{eq:pij} into

\begin{equation}
p_{ij}=\frac{zs_is_j}{1+zs_is_j};
\end{equation}
the only parameter $z$ is, then, determined by maximizing the log-likelihood $\ln P(\mathbf{A}|z)=\sum_{i=1}^N\sum_{j(>i)}[a_{ij}\ln p_{ij}+(1-a_{ij})\ln(1-p_{ij})]$, a recipe implying that the only equation to be solved reads

\begin{equation}\label{eq:ML}
\langle L\rangle=\sum_{i=1}^N\sum_{j(>i)}p_{ij}=\sum_{i=1}^N\sum_{j(>i)}\frac{zs_is_j}{1+zs_is_j}=L^*,
\end{equation}
where $L^*$ and $z^*_\text{ML}$ respectively indicate the empirical value of the total number of links and the related parameter estimation.

Instantiating the expression in eq.~\ref{eq:q} with the one defining the dcGM leads to

\begin{align}\label{eq:qq}
q_{ij}^{t+1}&=\int_0^{+\infty}\left(\frac{zs_i^{t+1}s_j^{t+1}}{1+zs_i^{t+1}s_j^{t+1}}\right)\frac{P(\mathbf{A}_t|z)\pi(z)}{P(\mathbf{A}_t)}dz,
\end{align}
where

\begin{equation}
P(\mathbf{A}_t|z)=\prod_{i=1}^{N_t}\prod_{j(>i)}(p_{ij}^t)^{a_{ij}^t}(1-p_{ij}^t)^{1-a_{ij}^t}
\end{equation}
and

\begin{equation}
p_{ij}^t=\frac{zs_i^ts_j^t}{1+zs_i^ts_j^t};
\end{equation}
let us stress that the dependence on the observed snapshot does not imply that the adjacency matrix at time $t$ must be entirely known: once the strength sequence is given, the binary information entering the likelihood is, in fact, summarized by the sufficient statistic $L_t$, i.e. the total number of links at time $t$ (see also Appendix~\hyperlink{AppC}{C}).

The rationale that led us to the BFM motivates us to determine $\pi(z)$ by adopting the recipe called \emph{empirical prior}: in words, one \emph{i)} estimates $z$ on each of the snapshots preceding the one under consideration; \emph{ii)} deduces the functional form of $\pi(z)$; \emph{iii)} plugs it into the expression of the snapshot to be reconstructed (see also Appendix~\hyperlink{AppC}{C}).

\begin{figure*}[t!]
\centering
\includegraphics[width=0.9\linewidth]{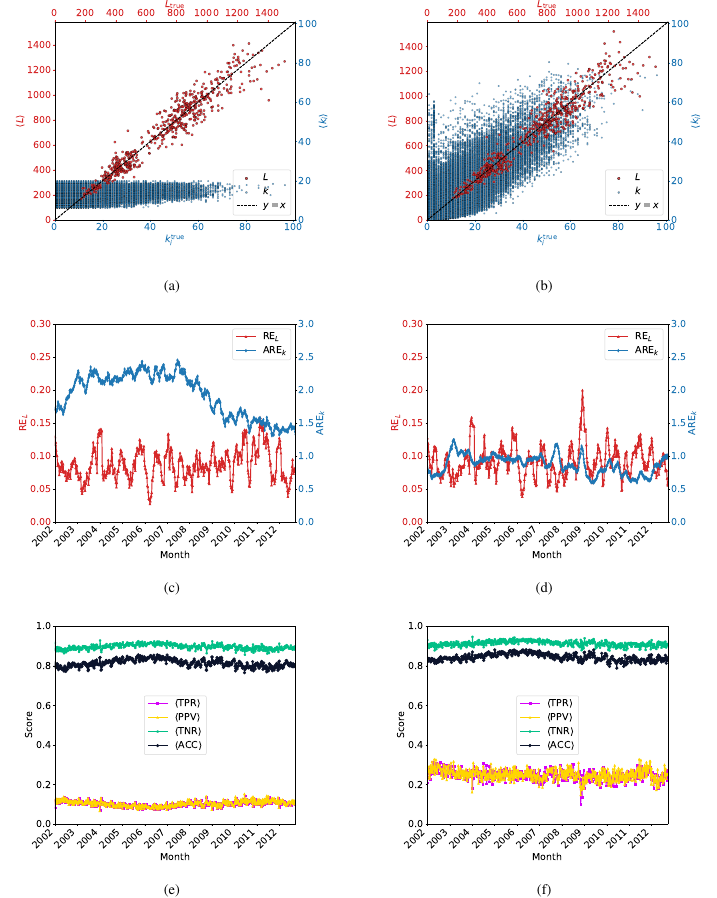}
\caption{\textbf{Out-of-sample reconstruction performance of the BERM and BFM.} Panels (a) and (b): empirical values of the total number of links (red) and node degrees (blue) scattered versus the predicted ones, pooled across the weeks constituting our dataset; the dashed line marks the identity.  Panels (c) and (d): evolution of the relative error on the total number of links (red) and the average relative error on the nodes degrees (blue), across the weeks constituting our dataset.  Panels (e) and (f): evolution of the $\langle\text{TPR}\rangle$, the $\langle\text{PPV}\rangle$, the $\langle\text{TNR}\rangle$ and the $\langle\text{ACC}\rangle$ across the weeks constituting our dataset. Panels (a), (c) and (e) refer to the BERM, while panels (b), (d) and (f) refer to the BFM: while both models recover the total number of links and achieve a large $\langle\text{ACC}\rangle$ score, driven by the large value of the $\langle\text{TNR}\rangle$, only the BFM is capable of recovering the degree sequence to an acceptable degree of accuracy - as well as more than doubling the other scores.}
\label{fig:1}
\end{figure*}

At this point, one needs to compute $q_{ij}^{t+1}$. A possible choice is that of evaluating the integrand in correspondence of $z^*_\text{MAP}$, i.e. the value that maximizes the log-posterior $\ln P(z|\mathbf{A}_t)=\ln P(\mathbf{A}_t|z)+\ln\pi(z)$, a recipe implying that the only equation to be solved reads

\begin{equation}
\frac{\partial\ln P(z|\mathbf{A}_t)}{\partial z}=\frac{\partial\ln P(\mathbf{A}_t|z)}{\partial z}+\frac{\partial\ln\pi(z)}{\partial z}=0
\end{equation}
or, more explicitly,

\begin{align}\label{eq:map}
\frac{1}{z}\left[L_t^*-\sum_{i=1}^{N_t}\sum_{j(>i)}\frac{zs_i^ts_j^t}{1+zs_i^ts_j^t}\right]+\frac{\partial\ln\pi(z)}{\partial z}=0;
\end{align}
although viable, this choice rests upon the assumption that the integrand is peaked around $z^*_\text{MAP}$; besides, it privileges the likelihood term over the prior - i.e. a system current realization over its history.

To avoid relying on this assumption, and fully account for uncertainty, one is forced to proceed numerically. To this aim, we move to the logarithmic coordinate $u=\ln z$ and employ the integration scheme named \emph{Gauss-Hermite quadrature} (see also the Methods section and Appendix~\hyperlink{AppD}{D}).

\subsection*{Out-of-sample network reconstruction\\ of the eMID dataset}

Let us, now, test our recipes on the transaction-level data constituting the overnight segment of the Electronic Market for Interbank Deposits (eMID), a screen-based market for unsecured deposits~\cite{marzi2026reproducing,macchiati2025spectral,IoriOvernightMoneyMarket2008,FingerFrickeLux2012}. Starting from directed, weighted trades (lender, borrower, notional amount), we aggregate transactions within ISO weeks, symmetrize exposures, and then binarize them to obtain weekly undirected snapshots (see also Appendix~\hyperlink{AppE}{E}).

The performance of our Bayesian ERGs has been tested on the following properties: the \emph{total number of links}, by comparing 

\begin{equation}
L_{t+1}=\sum_{i=1}^N\sum_{j(>i)}a_{ij}^{t+1}
\end{equation}
with $\langle L_{t+1}\rangle$; the \emph{degree of each node}, by comparing

\begin{equation}
k_i^{t+1}=\sum_{j(\neq i)}a_{ij}^{t+1},\quad\:\forall\:i
\end{equation}
with $\langle k_i^{t+1}\rangle$, $\forall\:i$; the \emph{true positive rate} (also known as \emph{recall} or \emph{sensitivity})

\begin{equation}\label{eq:tpr}
\langle\text{TPR}_{t+1}\rangle=\frac{\langle\text{TP}_{t+1}\rangle}{L_{t+1}},
\end{equation}
the \emph{positive predictive value} (also known as \emph{precision})

\begin{equation}
\langle\text{PPV}_{t+1}\rangle=\frac{\langle\text{TP}_{t+1}\rangle}{\langle L_{t+1}\rangle},
\end{equation}
the \emph{true negative rate} (also known as \emph{specificity})

\begin{equation}
\langle\text{TNR}_{t+1}\rangle=\frac{\langle\text{TN}_{t+1}\rangle}{V_{t+1}-L_{t+1}}
\end{equation}
and the \emph{accuracy}

\begin{equation}
\langle\text{ACC}_{t+1}\rangle=\frac{\langle\text{TP}_{t+1}\rangle+\langle\text{TN}_{t+1}\rangle}{V_{t+1}},
\end{equation}
quantifying the ability of an algorithm in capturing both the number of true positives and the number of true negatives.

We have also approached the problem of reconstructing a network at time $t+1$, on the basis of the information at time $t$, from the perspective of link prediction (although the related problem of spurious links detection can be approached within the same framework as well, we have ignored it, here): to this aim, we have considered the top $\langle L_{t+1}\rangle$ links, i.e. the links characterized by the $\langle L_{t+1}\rangle$ largest probability coefficients, and checked their position via the \emph{Jaccard Index}, reading

\begin{equation}
\text{JI}_{t+1}=\frac{\lvert E^\text{pred}_{t+1}\cap E^\text{true}_{t+1}\rvert}{\lvert E^\text{pred}_{t+1}\cup E^\text{true}_{t+1}\rvert},
\end{equation}
where $E^{\mathrm{true}}$ is the ground-truth edge set, with $\lvert E^{\mathrm{true}}\rvert=L_{t+1}$. We have also considered the Area Under the Receiver Operating Curve (AUROC), defined as the area under the curve obtained upon scattering the index

\begin{equation}
\text{TPR}_{t+1}=\frac{\lvert E^\text{pred}_{t+1}\cap E^\text{true}_{t+1}\rvert}{\lvert E^\text{true}_{t+1}\rvert}
\end{equation}
(notice that such an index is nothing but the deterministic version of the one defined in eq.~\ref{eq:tpr}) versus the index

\begin{figure*}[t!]
\centering
\includegraphics[width=\linewidth]{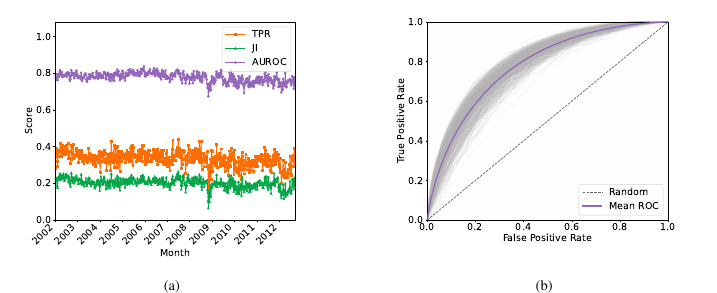}
\caption{\textbf{Ranking-based performance of the BFM.} Panel (a): evolution of the TPR, the JI and AUROC across the weeks constituting our dataset. The markers report pointwise weekly values; no error bars are shown. Panel (b): ROC curves for all snapshots. The purple one represents the average ROC, obtained by interpolating each snapshot-specific ROC on a common grid of FPR values and averaging the corresponding TPR values. The dashed diagonal line represents the random-classifier benchmark. These ranking-based diagnostics are meaningful only for the BFM, that induces a non-trivial ordering of candidate links.}
\label{fig:2}
\end{figure*}

\begin{equation}
\text{FPR}_{t+1}=\frac{\lvert E^\text{pred}_{t+1}\cap\overline{E^\text{true}_{t+1}}\rvert}{\lvert\overline{E^\text{true}_{t+1}}\rvert}
\end{equation}
(where $\overline{E^\text{true}_{t+1}}$ is the complementary set of $E^\text{true}_{t+1}$, with $\lvert\overline{E^\text{true}_{t+1}}\rvert=V_{t+1}-L_{t+1}$), as the list of links ranked in decreasing order of the chosen score is gone through: the AUROC quantifies the extent to which a given link prediction algorithm performs better than a random one - i.e. one that flips a coin to classify each non-observed link as either non-existent or missing.

Let us stress that the evaluation protocol above addresses two different aspects of the inference task, reflecting (the difference between) the reconstruction and the prediction tasks described in the Introduction: although the explicit sampling of individual network realizations is avoided in both cases, the quantities evaluated in a probabilistic fashion address the problem of reproducing the either global or local \emph{number of links}, while the ones evaluated in a deterministic fashion address the problem of reproducing their \emph{position}.

As a first analysis, we have considered the entire dataset, spanning the years $1999$–$2012$, at the weekly time scale: the first three years have been used for the initial calibration of the prior; from $2002$ on, we have performed the analysis over the remaining years, letting the prior be updated over a rolling window of three years. As all banks observed at each date are retained, here we have worked with an `unbalanced' panel of nodes - i.e. their number is allowed to vary from snapshot to snapshot. For what concerns the BERM, the empirical prior on $p^*$, across the initial calibration window, is well described by a beta distribution whose parameters read $\alpha\simeq58.261$ and $\beta\simeq541.543$ (see fig.~\ref{fig:appb2}). For what concerns the BFM, instead, the distribution of the $z^*$ values across the same calibration window is well described by the Gamma distribution $\pi(z)=z^{\kappa-1}e^{-z/\theta}/\Gamma(\kappa)\theta^\kappa$ whose parameters read $\kappa\simeq27.538$ and $\theta\simeq0.007$ - and the Kolmogorov-Smirnov test~\cite{Massey1951KSTest} confirms it to be the distribution with the smallest number of parameters (only two) not rejected as a plausible parent distribution (see fig.~\ref{fig:appc2}). In both cases, the prior is updated by adding each, new, weekly value to it: from this perspective, we register no substantial difference between enriching it with ML or MAP estimates (see also the Methods section).

As fig.~\ref{fig:1} shows, both the BERM and the BFM are capable of recovering the total number of links. When coming to the degrees, instead, the BFM recovers their heterogeneity to a substantially better extent: more quantitatively, \emph{i)} the relative error $\text{RE}_L=|L-\langle L\rangle|/L$ amounts, on average, to $\simeq0.1$, under both the BFM and the BERM; \emph{ii)} the average relative error $\text{ARE}_k=\sum_{i=1}^N(|k_i-\langle k_i\rangle|/k_i)/N$ is smaller under the BFM than under the BERM. Let us explicitly notice that, under the BFM, a deviation from the identity is visible for a limited set of nodes with very small degree. Such a mismatch is likely imputable to the underlying fitness ansatz, proxying the degree-controlling Lagrange multipliers with the node strengths: as observed in similar circumstances, this approximation works better for intermediate and large degrees than for small ones~\cite{CimiModel2015,Mazzarisi2017LimitedInformation,cimini2021reconstructing}.

\begin{figure*}[t!]
\centering
\includegraphics[width=\linewidth]{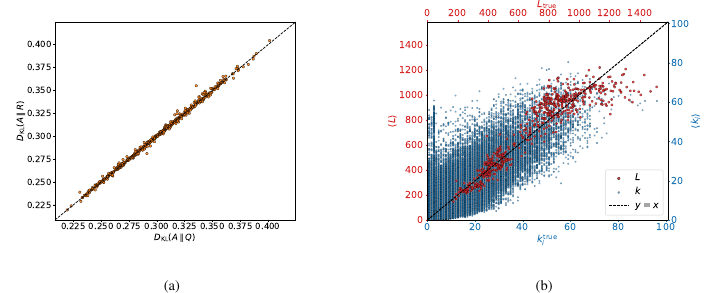}
\caption{\textbf{Comparison between direct and self-sustained reconstruction.} Panel (a): values of the Kullback-Leibler divergence between $\mathbf{A}$ and its ensemble average $\mathbf{Q}$ scattered versus the values of the Kullback-Leibler divergence between $\mathbf{A}$ and its `self-sustained' inferred version $\mathbf{R}$, pooled across the weeks constituting our dataset. Panel (b): values of the total number of links (red) and node degrees (blue) predicted by employing $\mathbf{Q}$ scattered versus the values predicted by employing $\mathbf{R}$, pooled across the weeks constituting our dataset. Both plots confirm that $\mathbf{Q}$ represents a reliable surrogate of $\mathbf{A}$ - in fact, so accurate to constitute a valid prior for subsequent inference.}
\label{fig:3}
\end{figure*}

Although both models achieve a large ACC, amounting, on average, to $\simeq0.80$, the BFM doubles the other scores, intended to quantify the ability of a model in recovering the position of connections - and not just that of the missing ones; the explanation of such a result lies in the evidence that the BERM cannot predict too dense configurations, although it is not capable of treating different pairs of nodes in a different way. Figure~\ref{fig:2} further refines the picture above, by plotting the TPR, the JI and the AUROC for the BFM: while the TPR amounts, on average, to $\simeq0.40$, the JI amounts, on average, to $\simeq0.20$ - but it should be noticed that the TPR of the Bayesian Fitness Model is twice the one achieved in~\cite{Parisi2018EntropyMissingLinks} by the Directed Binary Configuration Model, implemented to carry out the in-sample version of the same exercise, on the same dataset. We explicitly stress that these ranking-based diagnostics are meaningful only for the BFM, whose heterogeneous set of probability coefficients induces a non-trivial ordering of the candidate links; the BERM, instead, would assign the same score to each pair, therefore inducing a `random' classification.

A simpler version of our analysis (carried out by considering a `balanced' panel of $73$ banks across the weeks of $2002$ - i.e. retaining only the nodes that appear in all snapshots of both the calibration and test period) is described in Appendix~\hyperlink{AppB}{B} and Appendix~\hyperlink{AppC}{C}.

\begin{figure*}[t!]
\centering
\includegraphics[width=\linewidth]{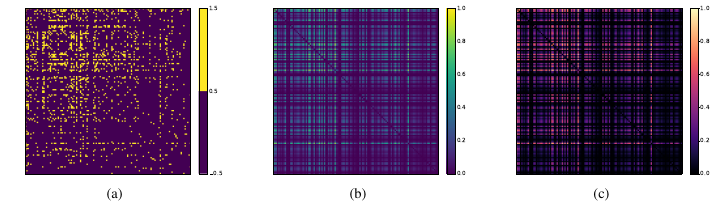}
\caption{\textbf{Observed adjacency and Bayesian reconstructions for a representative snapshot.} Panel (a): empirical adjacency matrix $\mathbf{A}_{t+1}$ corresponding to the week $\#20$ of the year $2007$. Panel (b): ensemble average of $\mathbf{A}_{t+1}$, i.e. $\mathbf{Q}_{t+1}$. Panel (c): `self-sustained', inferred version of $\mathbf{A}_{t+1}$, i.e. $\mathbf{R}_{t+1}$. While $\mathbf{Q}_{t+1}$ needs the information provided by $\mathbf{A}_t$, $\mathbf{R}_{t+1}$ `only' needs the information provided by $\mathbf{Q}_t$, i.e. an estimate of $\mathbf{A}_t$. More quantitatively, $2\sum_{i=1}^N\sum_{j(>i)}|q_{ij}-r_{ij}|/N(N-1)\simeq0.006$ and $2\sum_{i=1}^N\sum_{j(>i)}|a_{ij}-q_{ij}|/N(N-1)\simeq0.152\simeq2\sum_{i=1}^N\sum_{j(>i)}|a_{ij}-r_{ij}|/N(N-1)$.}
\label{fig:4}
\end{figure*}

\subsection*{Self-sustained inference of the eMID dataset}

The previous exercise was carried out by employing $\mathbf{A}_0$ (where $t=0$ indicates the last week of $2001$) to predict $\mathbf{A}_1$, $\mathbf{A}_1$ to predict $\mathbf{A}_2$ and so on. Let us, now, test the capability of our inference procedure to \emph{self-sustain} itself, i.e. to operate recursively without accessing any other adjacency matrix beyond the one employed for initialization. In this regime, both the prior on the parameter and the network representation are updated in a fully Bayesian fashion: the empirical prior keeps rolling over a three-year window but each new parameter value is estimated from inferred quantities and the matrix $\mathbf{Q}_t$ replaces $\mathbf{A}_t$ in the predictive step.

More specifically, we still consider the entire dataset by letting the prior roll over a time window of three years but \emph{i)} determine each, new, weekly value of the parameter by solving eq.~\ref{eq:map} where $L_t^*$ is replaced by $\langle L_t\rangle$, i.e. an estimate itself; \emph{ii)} substitute the generic coefficient $q_{ij}^{t+1}$ with the expression

\begin{equation}
r_{ij}^{t+1}=\int_0^{+\infty}\left(\frac{zs_i^{t+1}s_j^{t+1}}{1+zs_i^{t+1}s_j^{t+1}}\right)\frac{P(\mathbf{Q}_t|z)\pi(z)}{P(\mathbf{Q}_t)}dz,
\end{equation}
where $\mathbf{A}_t$ is replaced by $\mathbf{Q}_t$, again an estimate itself: stated otherwise, we employ $\mathbf{A}_0$ to predict $\mathbf{Q}_1$, $\mathbf{Q}_1$ to predict $\mathbf{Q}_2$ and so on.

As fig.~\ref{fig:3} shows, such a `self-sustained' inference highlights patterns that are very similar to those in fig.~\ref{fig:1}, an evidence revealing the accuracy of our algorithm to carry out an out-of-sample network reconstruction in presence of very little information (see also Appendix~\hyperlink{AppF}{F}). To provide a more quantitative assessment of such an agreement, let us scatter the values of the Kullback-Leibler divergence between the adjacency matrix $\mathbf{A}=\{a_{ij}\}$ and its ensemble average $\mathbf{Q}=\{q_{ij}\}$ at time $t$, i.e.

\begin{equation}
D_\text{KL}(\mathbf{A}_t||\mathbf{Q}_t)=\sum_{i=1}^N\sum_{j(>i)}a_{ij}^t\ln\left(\frac{a_{ij}^t}{q_{ij}^t}\right),
\end{equation}
versus the values of the Kullback-Leibler divergence between the adjacency matrix $\mathbf{A}=\{a_{ij}\}$ and its `self-sustained', inferred version $\mathbf{R}=\{r_{ij}\}$ at time $t$, i.e.,

\begin{equation}
D_\text{KL}(\mathbf{A}_t||\mathbf{R}_t)=\sum_{i=1}^N\sum_{j(>i)}a_{ij}^t\ln\left(\frac{a_{ij}^t}{r_{ij}^t}\right);
\end{equation}
let us stress that the quantities above are not intended to represent Kullback-Leibler divergences between joint distributions: they are, in fact, \emph{defined} in an element-wise fashion to compare all empirical entries with their reconstructed counterparts at once - in the same spirit of the discrepancy measure employed in~\cite{Young2020BayesianUnreliable}.

As evident from the figure, the two series of weekly values closely resemble each other. An even more explicit representation of such an agreement is provided in fig.~\ref{fig:4}, depicting the empirical adjacency matrix of the week $\#20$ of the year $2007$, its ensemble average $\mathbf{Q}$ and its `self-sustained' inferred version $\mathbf{R}$.

\subsection*{Out-of-sample versus in-sample reconstruction}

\begin{figure*}[t!]
\centering
\includegraphics[width=\linewidth]{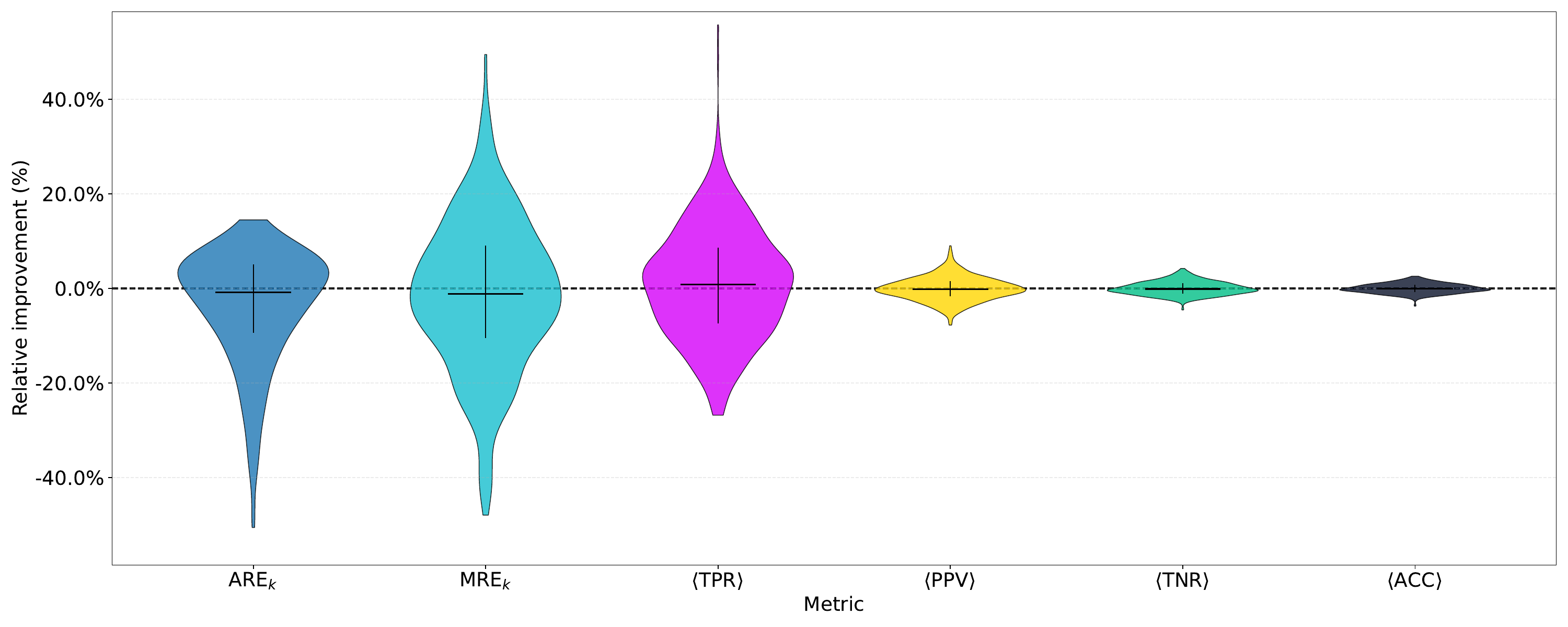}
\caption{\textbf{Metric-specific improvement of the `self-sustained' BFM over the in-sample dcGM.} For each snapshot and metric $m$, we define the score $I_m$ in two different ways, i.e. as $I_m=(m_{\text{dcGM}}-m_{\text{Bayes}})/m_{\text{dcGM}}$ for the $\text{ARE}_k$ and the $\text{MRE}_k$ and as $I_m=(m_{\text{Bayes}}-m_{\text{dcGM}})/m_{\text{dcGM}}$ for the $\langle\text{TPR}\rangle$, the $\langle\text{PPV}\rangle$, the $\langle\text{TNR}\rangle$ and the $\langle\text{ACC}\rangle$: in both cases, values above the $0\%$ dashed line indicate that the Bayesian predictor performs better than the (in-sample) dcGM. Each violin plot summarizes the distribution of the improvement, showing that our fully predictive procedure frequently matches, and sometimes exceeds, the in-sample reconstruction calibrated by taking $L$ as input at each time step.}
\label{fig:5}
\end{figure*}

A particularly stringent test of our predictive scheme is obtained by contrasting it with an in-sample reconstruction that is allowed to use the quantity that our method must, instead, infer. More specifically, we compare the set of $r_{ij}$s with the set of probability coefficients returned by the dcGM, calibrated by employing the information about the total number of links per (weekly) snapshot. Despite the `informational advantage' of the dcGM, the predictive performances of the two models remain remarkably close: as fig.~\ref{fig:5} shows, upon averaging over $561$ weekly snapshots we obtain $\overline{\text{TPR}}_\text{Bayes}=0.2498$ versus $\overline{\text{TPR}}_\text{dcGM}=0.2494$, with the Bayesian predictor outperforming the dcGM on the $\simeq54\%$ of snapshots. For what concerns the remaining scores, we have $\overline{\text{PPV}}_\text{Bayes}=0.2495$ versus $\overline{\text{PPV}}_\text{dcGM}=0.2494$, $\overline{\text{TNR}}_\text{Bayes}=0.9124$ versus $\overline{\text{TNR}}_\text{dcGM}=0.9122$ and $\overline{\text{ACC}}_\text{Bayes}=0.8429$ versus $\overline{\text{ACC}}_\text{dcGM}=0.8432$; degree-level errors, instead, remain slightly smaller for the dcGM, with $\overline{\text{ARE}}_k^\text{Bayes}=0.8999$ versus $\overline{\text{ARE}}_k^\text{dcGM}=0.8624$ and $\text{MRE}_k^\text{Bayes}=14.6836$ versus $\text{MRE}_k^\text{dcGM}=14.6267$ (with $\text{MRE}_k=\max_{i=1}^N\{|k_i-\langle k_i\rangle|/k_i\}$), although the Bayesian predictor is still better for a non-negligible fraction of weeks, i.e. on the $\simeq46\%$ of the snapshots.

\section*{DISCUSSION}

In the economic and financial domain, reconstruction methods based on Shannon entropy maximization have been widely studied from a frequentist perspective~\cite{ReconstructionMethods2018,cimini2019statistical,cimini2021reconstructing}. Within such a framework, fitness-based models provide a powerful approach to reconstructing the structure of complex networks from partial information: among them, the dcGM offers a minimal, yet effective, parametrization for estimating link probabilities that combines node-level attributes with a global parameter.

The dcGM accurately captures many structural features of real-world networks but relies on a point-wise, maximum-of-the-likelihood estimation procedure that offers no principled way to incorporate temporal dependencies: in other words, the dcGM can be employed to carry out an \emph{in-sample} network reconstruction.

Related predictive tasks have been studied within the literature on temporal networks: a first example is provided by Temporal Exponential Random Graphs that, however, can be employed to predict future edges only in case a sequence of empirical networks with the same set of nodes is available for parameter estimation~\cite{Hanneke2010,Krivitsky2014}; other examples are provided by machine learning and embedding-based approaches that, however, operate under a richer informational regime than the one considered here, requiring fully observable past snapshots, stable representations of features, \emph{ad-hoc} mechanisms to handle nodes entering or leaving the system, validation procedures for tuning hyperparameters~\cite{lu_link_2011,Zhou_2021,santucci_missing_2026}.

Against this background, building on the approach developed in~\cite{Peixoto2018PRXErrors,Peixoto2025PRXMDL,Young2020BayesianUnreliable}, our contribution addresses the problem of carrying out an \emph{out-of-sample} network reconstruction, by presenting a fully Bayesian framework to infer a network structure from a previous observation. As our tests reveal, our reconstruction procedure is capable of `self-sustaining' itself, each predicted configuration becoming a reliable prior for inference at the subsequent step. In practice, after the lastly observed snapshot is used, no additional topological information is fed into the algorithm and the reconstruction is propagated forward in time solely through the posterior predictive distribution: such an experiment probes the information-retention capability of the model, as prediction errors are allowed to accumulate while our procedure attempts to reconstruct future configurations at increasingly large temporal distances from the lastly observed topology.

\section*{METHODS}

\subsection*{Data preprocessing}

We analyze the binary undirected representation of eMID across the years $1999$-$2012$: weekly snapshots are built from all transactions settled within ISO weeks; self-loops and multi-edges are discarded. For the unbalanced specification, the empirical prior is calibrated on the years $1999$-$2001$ while the out-of-sample analysis is performed across the years $2002$-$2012$, retaining all the banks that are active in each week. For the balanced specification, discussed in Appendix~\hyperlink{AppB}{B} (BERM) and Appendix~\hyperlink{AppC}{C} (BFM), the empirical prior is calibrated on the years $1999$-$2001$ and the node set is restricted to the banks that are active in all weekly snapshots from $1999$ to $2002$; the out-of-sample analysis is, then, performed only over the year $2002$. A complete description of the raw records, weekly aggregation, symmetrization and binarization steps is provided in Appendix~\hyperlink{AppE}{E}.

\subsection*{Prior calibration}

Within each rolling window of three years, we compute a snapshot-specific point estimate for each parameter and, then, fit the empirical prior on the resulting set of values. Such a set is augmented via the jack-knife procedure, i.e. by removing one trading day at a time from each weekly aggregation and repeating the point-estimation step.

For what concerns the BERM, the ML estimate reads $p^*=L^*/V^*$; for what concerns the BFM, instead, the ML estimate is obtained by solving eq.~\ref{eq:ML}: in practice, we solve the corresponding one-dimensional problem via Newton iterations with backtracking, using a tolerance of $10^{-8}$ and a maximum number of $80$ iterations. The corresponding fits on the initial calibration window are shown in Appendix~\hyperlink{AppB}{B} for the BERM and in Appendix~\hyperlink{AppC}{C} for the BFM.

\subsection*{Evaluation of the posterior predictive distribution}

Adopting the Beta distribution as a (conjugate) prior allows us to treat the posterior predictive distribution of the BERM analytically.

For what concerns the BFM, instead, the posterior predictive distribution must be evaluated numerically. To this aim, we pose $u=\ln z$ and perform the calculation in such a logarithmic coordinate, which maps the original domain $(0,+\infty)$ onto $\mathbb{R}$ and prevents numerical overflow when $z$ becomes large.

A first integration scheme is the one named \emph{Gauss-Hermite quadrature}~\cite{AbramowitzStegun1964}: after having `gaussianized' the integrand and evaluated it in correspondence of the posterior mode, the predictive probability reduces to a weighted sum over $K$ quadrature nodes; for our experiments, we have used $K=25$.

An alternative integration scheme is the one named \emph{slice sampling}~\cite{Neal2003SliceSampling}: it samples the log-posterior $M$ times without requiring the computation of the normalization constant and providing an estimator that converges to the true distribution as $M$ increases; for our experiments, $M=3000$ values are sampled after a burn-in phase of $600$ steps, thus yielding the estimator

\begin{equation}
q_{ij}^{t+1}\simeq\frac{1}{M}\sum_{m=1}^{M}\frac{e^{u^{(m)}}s_i^{t+1}s_j^{t+1}}{1+e^{u^{(m)}}s_i^{t+1}s_j^{t+1}}.
\end{equation}

While the snapshot-specific computational cost of the first scheme scales as $O(KN^2)$, the one of the second scheme scales as $O(MN^2)$: as the Gauss-Hermite quadrature one is computationally cheaper, it is employed as the default method (see also Appendix~\hyperlink{AppD}{D}).

\section*{DATA AVAILABILITY}

Raw eMID data are subject to restrictions, hence not publicly available. Researchers can access them upon request.

\section*{CODE AVAILABILITY}

The Python package named \texttt{OR4CLE} (\textit{Out-of-sample bayesian Reconstruction 4 CompLex nEtworks}), implementing the algorithms described in the main text, is available on PyPI and at the URL \url{https://github.com/mattiamarzi/OR4CLE}.

\bibliography{references}

\section*{ACKNOWLEDGMENTS}

MM and TS acknowledge support from the project ‘SoBigData.it - Strengthening the Italian RI for Social Mining and Big Data Analytics’ - IR0000013 - CUP B53C22001760006, financed by European Union - Next Generation EU - National Recovery and Resilience Plan (Piano Nazionale di Ripresa e Resilienza, PNRR) - M4C2 I.3.1; TS acknowledges support from the projects `RE-Net: Reconstructing economic networks: from physics to machine learning and back' - 2022MTBB22, Funded by the European Union Next Generation EU, PNRR Mission 4 Component 2 Investment 1.1, CUP: D53D23002330006; `C2T - From Crises to Theory: towards a science of resilience and recovery for economic and financial systems' - P2022E93B8, Funded by the European Union Next Generation EU, PNRR Mission 4 Component 2 Investment 1.1, CUP: D53D23019330001.

\section*{AUTHOR CONTRIBUTIONS}

Study conception and design: MM, TS. Analysis and interpretation of results: MM, TS. Draft manuscript preparation: MM, TS.

\section*{COMPETING INTERESTS} 

The authors declare no competing interests.

\clearpage
\onecolumngrid

\appendix

\counterwithin*{figure}{section}
\stepcounter{section}
\renewcommand{\thefigure}{A.\arabic{figure}}

\section*{APPENDIX A.\\More on the posterior predictive distribution}
\hypertarget{AppA}{}

Since

\begin{equation}
P(\mathbf{A}_{t+1}|\mathbf{A}_t)=\int_0^{+\infty}P(\mathbf{A}_{t+1}|z)P(z|\mathbf{A}_t)dz=\int_0^{+\infty}\prod_{i=1}^N\prod_{j(>i)}\left(p_{ij}^{t+1}\right)^{a_{ij}^{t+1}}\left(1-p_{ij}^{t+1}\right)^{1-a_{ij}^{t+1}}P(z|\mathbf{A}_t)dz
\end{equation}
and

\begin{align}
\sum_{a_{ij}^{t+1}=0}^1\left(p_{ij}^{t+1}\right)^{a_{ij}^{t+1}}\left(1-p_{ij}^{t+1}\right)^{1-a_{ij}^{t+1}}=1,
\end{align}
one finds that

\begin{align}
\langle a_{ij}^{t+1}|\mathbf{A}_t\rangle&=\sum_{\mathbf{A}_{t+1}}a_{ij}^{t+1}P(\mathbf{A}_{t+1}|\mathbf{A}_t)\nonumber\\
&=\sum_{a_{ij}^{t+1}}\sum_{\mathbf{A}_{t+1}\setminus a_{ij}^{t+1}}a_{ij}^{t+1}P(\mathbf{A}_{t+1}|\mathbf{A}_t)\nonumber\\
&=0\cdot P(a_{ij}^{t+1}=0|\mathbf{A}_t)+1\cdot P(a_{ij}^{t+1}=1|\mathbf{A}_t)\nonumber\\
&=P(a_{ij}^{t+1}=1|\mathbf{A}_t).
\end{align}

\clearpage

\counterwithin*{figure}{section}
\stepcounter{section}
\renewcommand{\thefigure}{B.\arabic{figure}}

\section*{APPENDIX B.\\The Bayesian Erd\"os-R\'enyi Model}
\hypertarget{AppB}{}

Let us, now, fully illustrate the calculations concerning the Erd\"os-R\'enyi Model. Since its classical formulation reads

\begin{equation}
P(\mathbf{A}|p)=p^{L(\mathbf{A})}(1-p)^{V-L(\mathbf{A})},
\end{equation}
with $V=N(N-1)/2$, enriching it with a conjugate prior amounts to considering the expression

\begin{align}
P(p|\mathbf{A})&=\frac{P(\mathbf{A}|p)\pi(p)}{\int_0^1P(\mathbf{A}|p)\pi(p)dp}=\frac{p^{L(\mathbf{A})+\alpha-1}(1-p)^{V-L(\mathbf{A})+\beta-1}}{\int_0^1p^{L(\mathbf{A})+\alpha-1}(1-p)^{V-L(\mathbf{A})+\beta-1}dp}=\frac{p^{L(\mathbf{A})+\alpha-1}(1-p)^{V-L(\mathbf{A})+\beta-1}}{\text{B}(L(\mathbf{A})+\alpha,V-L(\mathbf{A})+\beta)},
\end{align}
coming from posing

\begin{align}
\pi(p)=\frac{p^{\alpha-1}(1-p)^{\beta-1}}{\text{B}(\alpha,\beta)},
\end{align}
where $\text{B}(\alpha,\beta)=\int_0^1p^{\alpha-1}(1-p)^{\beta-1}dp$ is the Beta function. We can, thus, write

\begin{align}
P(L_{t+1}=k|\mathbf{A}_t)&=\int_0^1P(L_{t+1}=k|p)P(p|\mathbf{A}_t)dp\nonumber\\
&=\int_0^1\binom{V_{t+1}}{k}p^k(1-p)^{V_{t+1}-k}P(p|\mathbf{A}_t)dp\nonumber\\
&=\int_0^1\binom{V_{t+1}}{k}p^k(1-p)^{V_{t+1}-k}\frac{P(\mathbf{A}_t|p)\pi(p)}{P(\mathbf{A}_t)}dp
\end{align}
and

\begin{figure*}[t!]
\centering
\includegraphics[width=0.9\linewidth]{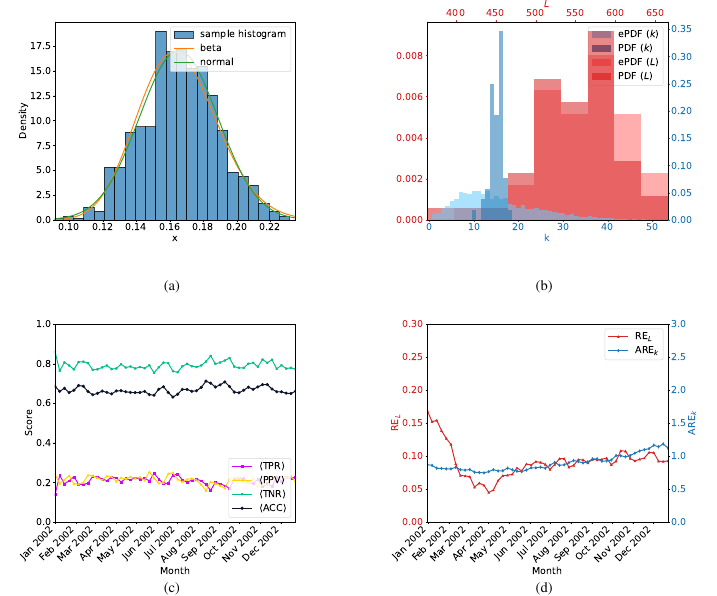}
\caption{\textbf{Performance of the BERM on the balanced snapshots.} Panel (a): histogram of the estimated $p^*$ values across the calibration set consisting of the weeks, augmented via the jack-knife method, constituting the years $1999$-$2001$ together with two fitted PDFs, i.e. a beta distribution (orange line) and a normal distribution (green line); both of them provide a reasonable fit and they are `saved' from the Kolmogorov-Smirnov test. Panel (b): empirical VS predicted PDF of the total number of links (red) and nodes degrees (blue), pooled across the weeks constituting the year $2002$. Panel (c): evolution of the $\langle\text{TPR}\rangle$, the $\langle\text{PPV}\rangle$, the $\langle\text{TNR}\rangle$ and the $\langle\text{ACC}\rangle$ across the weeks constituing the year $2002$. Panel (d): evolution of the relative error on the total number of links (red) and the average relative error on the node degrees (blue), across the weeks constituting the year $2002$.}
\label{fig:appb1}
\end{figure*}

\begin{align}
P(\mathbf{A}_t)=\int_0^1P(\mathbf{A}_t|p)\pi(p)dp=\int_0^1p^{L_t+\alpha-1}(1-p)^{V_t-L_t+\beta-1}dp=\text{B}(L_t+\alpha,V_t-L_t+\beta),
\end{align}
where the explicit dependence of the expression above on $\mathbf{A}_t$ has been dropped. Putting everything together, one gets

\begin{align}
P(L_{t+1}=k|\mathbf{A}_t)=\binom{V_{t+1}}{k}\frac{\int_0^1p^{k+L_t+\alpha-1}(1-p)^{V_{t+1}+V_t-k-L_t+\beta-1}dp}{\int_0^1p^{L_t+\alpha-1}(1-p)^{V_t-L_t+\beta-1}dp}=\binom{V_{t+1}}{k}\frac{\text{B}(k+L_t+\alpha,V_{t+1}+V_t-k-L_t+\beta)}{\text{B}(L_t+\alpha,V_t-L_t+\beta)},
\end{align}
i.e. the total number of links at time $t+1$, conditional to the observation of $\mathbf{A}_t$, obeys the beta-binomial distribution $\text{BetaBin}(V_{t+1},L_t+\alpha,V_t-L_t+\beta)$. Since

\begin{align}
\langle L_{t+1}|\mathbf{A}_t\rangle=\sum_{k=0}^{V_{t+1}}kP(L_{t+1}=k|\mathbf{A}_t)=\sum_{k=0}^{V_{t+1}}k\int_0^1P(L_{t+1}=k|p)P(p|\mathbf{A}_t)dp,
\end{align}
swapping the operations of sum and integration leads to

\begin{align}
\langle L_{t+1}|\mathbf{A}_t\rangle&=\int_0^1\sum_{k=0}^{V_{t+1}}kP(L_{t+1}=k|p)P(p|\mathbf{A}_t)dp
\end{align}
and recognizing in $P(L_{t+1}=k|p)$ a binomial distribution further leads to

\begin{align}
\langle L_{t+1}|\mathbf{A}_t\rangle&=\int_0^1\sum_{k=0}^{V_{t+1}}k\binom{V_{t+1}}{k}p^k(1-p)^{V_{t+1}-k}\frac{P(\mathbf{A}_t|p)\pi(p)}{P(\mathbf{A}_t)}dp\nonumber\\
&=\int_0^1V_{t+1}p\frac{P(\mathbf{A}_t|p)\pi(p)}{P(\mathbf{A}_t)}dp\nonumber\\
&=V_{t+1}\int_0^1p\frac{P(\mathbf{A}_t|p)\pi(p)}{P(\mathbf{A}_t)}dp.
\end{align}

Comparing the expression above with equation from the main text
\begin{align}
q_{ij}^{t+1}&=P(a_{ij}^{t+1}=1|\mathbf{A}_t)=\int_0^{+\infty}p_{ij}^{t+1}(z)\frac{P(\mathbf{A}_t|z)\pi(z)}{P(\mathbf{A}_t)}dz,
\end{align}
leads us to name the second factor $q^{t+1}$: hence,

\begin{align}
q^{t+1}=\int_0^1p\frac{P(\mathbf{A}_t|p)\pi(p)}{P(\mathbf{A}_t)}dp=\frac{\int_0^1p^{1+L_t+\alpha-1}(1-p)^{V_t-L_t+\beta-1}dp}{\int_0^1p^{L_t+\alpha-1}(1-p)^{V_t-L_t+\beta-1}dp}=\frac{\text{B}(1+L_t+\alpha,V_t-L_t+\beta)}{\text{B}(L_t+\alpha,V_t-L_t+\beta)}=\frac{L_t+\alpha}{V_t+\alpha+\beta},
\end{align}
where we have employed the identity

\begin{align}
\frac{\text{B}(x+1,y)}{\text{B}(x,y)}=\frac{\Gamma(x+1)\Gamma(y)}{\Gamma(x+y+1)}\cdot\frac{\Gamma(x+y)}{\Gamma(x)\Gamma(y)}=\frac{x\Gamma(x)\Gamma(y)}{(x+y)\Gamma(x+y)}\cdot\frac{\Gamma(x+y)}{\Gamma(x)\Gamma(y)}=\frac{x}{x+y}.
\end{align}

The performance of the model in reproducing the chosen empirical quantities across the weeks of the year $2002$ is shown in fig.~\ref{fig:appb1}. As their number amounts to $\simeq150$, we have implemented the jack-knife method to augment the sample: more specifically, we have removed each day of each week at a time, hence producing $7$ weeks of $6$ days each out of $1$ week of $7$ days; upon doing so, we have moved from $\simeq150$ snapshots to $\simeq1000$ snapshots. Here, we have focused on a balanced panel of $N=73$ banks.

\begin{figure*}[t!]
\centering
\includegraphics[width=0.9\linewidth]{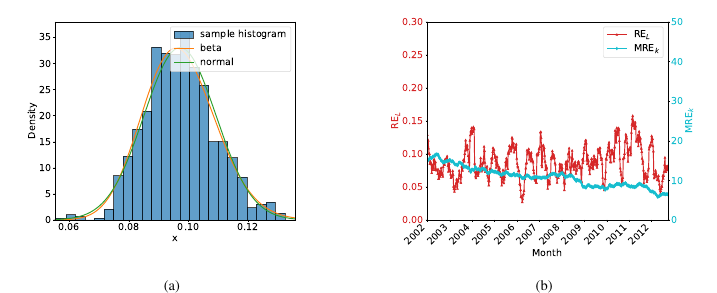}
\caption{\textbf{Performance of the BERM on the unbalanced snapshots.} Panel (a): histogram of the estimated $p^*$ values across the calibration set consisting of the weeks (augmented via the jack-knife method) constituting the years $1999$-$2001$ together with two, fitted PDFs, i.e. a beta distribution (orange line) and a normal distribution (green line); both of them provide a reasonable fit, as confirmed by the Kolmogorov-Smirnov test. Panel (b): evolution of the relative error on the total number of links (red) and the maximum relative error on the node degrees (cyan), across the weeks constituting our dataset.}
\label{fig:appb2}
\end{figure*}

The performance of the model in reproducing the chosen empirical quantities across the weeks constituting our dataset is, instead, shown in fig.~\ref{fig:appb2}. Its right panel deserves to be commented more. First, let us notice that the $\text{MRE}_k$ decreases: this seems to suggest that the network topology is becoming increasingly compatible with such a model - although overly simple. Second, let us notice the bump in correspondence of the year $2008$: this suggests that the crisis reveals itself as an event challenging the model adopted to describe the system under consideration, letting the error associated to the estimates of the degrees rise. Overall, however, the two evidences above confirm the results obtained in analogous papers, i.e. that financial systems somehow `lose' structure after a crisis.

\clearpage

\counterwithin*{figure}{section}
\stepcounter{section}
\renewcommand{\thefigure}{C.\arabic{figure}}

\section*{APPENDIX C.\\The Bayesian Fitness Model}
\hypertarget{AppC}{}

Let us, now, fully illustrate the calculations concerning the Fitness Model. Overall,

\begin{align}
P(\mathbf{A}_{t+1}|\mathbf{A}_t)&=\int_0^{+\infty}P(\mathbf{A}_{t+1}|z)P(z|\mathbf{A}_t)dz\nonumber\\
&=\int_0^{+\infty}\frac{P(\mathbf{A}_{t+1}|z)P(\mathbf{A}_t|z)\pi(z)}{P(\mathbf{A}_t)}dz\nonumber\\
&=\frac{\int_0^{+\infty}\left[\prod_{i=1}^N\prod_{j(>i)}\left(p_{ij}^{t+1}\right)^{a_{ij}^{t+1}}\left(1-p_{ij}^{t+1}\right)^{1-a_{ij}^{t+1}}\right]\left[\prod_{i=1}^N\prod_{j(>i)}\left(p_{ij}^t\right)^{a_{ij}^t}\left(1-p_{ij}^t\right)^{1-a_{ij}^t}\right]\pi(z)dz}{\int_0^{+\infty}\left[\prod_{i=1}^N\prod_{j(>i)}\left(p_{ij}^t\right)^{a_{ij}^t}\left(1-p_{ij}^t\right)^{1-a_{ij}^t}\right]\pi(z)dz}\nonumber\\
&=\frac{\int_0^{+\infty}\left[\prod_{i=1}^N\prod_{j(>i)}\left(\frac{p_{ij}^{t+1}}{1-p_{ij}^{t+1}}\right)^{a_{ij}^{t+1}}\left(1-p_{ij}^{t+1}\right)\right]\left[\prod_{i=1}^N\prod_{j(>i)}\left(\frac{p_{ij}^t}{1-p_{ij}^t}\right)^{a_{ij}^t}\left(1-p_{ij}^t\right)\right]\pi(z)dz}{\int_0^{+\infty}\left[\prod_{i=1}^N\prod_{j(>i)}\left(\frac{p_{ij}^t}{1-p_{ij}^t}\right)^{a_{ij}^t}\left(1-p_{ij}^t\right)\right]\pi(z)dz}\nonumber\\
&=\frac{\int_0^{+\infty}\left[\prod_{i=1}^N\prod_{j(>i)}\frac{\left(zs_i^{t+1}s_j^{t+1}\right)^{a_{ij}^{t+1}}}{1+zs_i^{t+1}s_j^{t+1}}\right]\left[\prod_{i=1}^N\prod_{j(>i)}\frac{\left(zs_i^ts_j^t\right)^{a_{ij}^t}}{1+zs_i^ts_j^t}\right]\pi(z)dz}{\int_0^{+\infty}\left[\prod_{i=1}^N\prod_{j(>i)}\frac{\left(zs_i^ts_j^t\right)^{a_{ij}^t}}{1+zs_i^ts_j^t}\right]\pi(z)dz};
\end{align}
in words, predictive reconstruction assumes a shared density scale across $t$ and $t+1$, together with conditional independence. The strengths, instead, are allowed to vary over time.

\begin{figure*}[t!]
\centering
\includegraphics[width=0.9\linewidth]{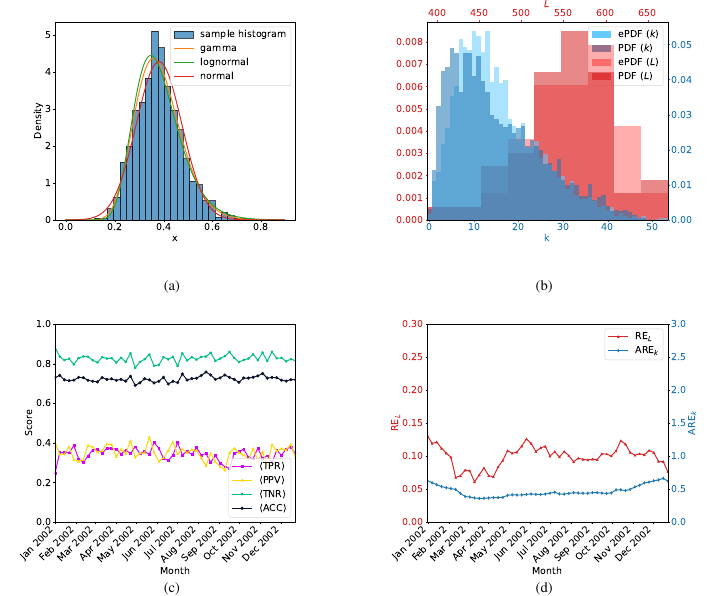}
\caption{\textbf{Performance of the BFM on the balanced snapshots.} Panel (a): histogram of the estimated $z^*$ values across the calibration set consisting of the weeks (augmented via the jack-knife method) constituting the years $1999$-$2001$ together with three, fitted PDFs, i.e. a gamma distribution (orange line), a normal distribution (red line) and a log-normal distribution (green line); although all of them provide a reasonable fit, the Kolmogorov-Smirnov test only `saves' the gamma distribution. Panel (b): empirical VS predicted PDF of the total number of links (red) and nodes degrees (blue), pooled across the weeks constituting the year $2002$. Panel (c): evolution of the $\langle\text{TPR}\rangle$, the $\langle\text{PPV}\rangle$, the $\langle\text{TNR}\rangle$ and the $\langle\text{ACC}\rangle$ across the weeks constituting the year $2002$. Panel (d): evolution of the relative error on the total number of links (red) and the average relative error on the node degrees (blue), across the weeks constituting the year $2002$.}
\label{fig:appc1}
\end{figure*}

Let us notice that the formulas above are valid to analyze a balanced panel of nodes ($N_t=N_{t+1}=N$); nothing, however, prevents us from considering an unbalanced one: in such a case,

\begin{align}
P(\mathbf{A}_{t+1}|\mathbf{A}_t)&=\frac{\int_0^{+\infty}\left[\prod_{i=1}^{N_{t+1}}\prod_{j(>i)}\frac{\left(zs_i^{t+1}s_j^{t+1}\right)^{a_{ij}^{t+1}}}{1+zs_i^{t+1}s_j^{t+1}}\right]\left[\prod_{i=1}^{N_t}\prod_{j(>i)}\frac{\left(zs_i^ts_j^t\right)^{a_{ij}^t}}{1+zs_i^ts_j^t}\right]\pi(z)dz}{\int_0^{+\infty}\left[\prod_{i=1}^{N_t}\prod_{j(>i)}\frac{\left(zs_i^ts_j^t\right)^{a_{ij}^t}}{1+zs_i^ts_j^t}\right]\pi(z)dz}.
\end{align}

\begin{figure*}[t!]
\centering
\includegraphics[width=0.9\linewidth]{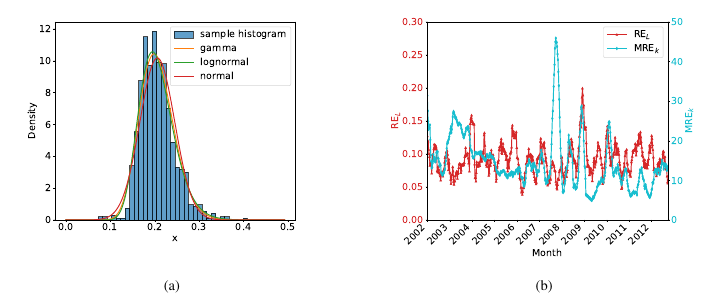}
\caption{\textbf{Performance of the BFM on the unbalanced snapshots.} Panel (a): histogram of the estimated $z^*$ values across the calibration set consisting of the weeks (augmented via the jack-knife method) constituting the years $1999$-$2001$ together with three, fitted PDFs, i.e. a gamma distribution (orange line), a normal distribution (red line) and a log-normal distribution (green line); although all of them provide a reasonable fit, the Kolmogorov-Smirnov test only `saves' the gamma distribution. Panel (b): evolution of the relative error on the total number of links (red) and the maximum relative error on the node degrees (blue), across the weeks constituting our dataset.}
\label{fig:appc2}
\end{figure*}

The performance of the model in reproducing the chosen empirical quantities across the weeks of the year $2002$ is shown in fig.~\ref{fig:appc1}. As their number amounts to $\simeq150$, we have implemented the jack-knife method to augment the sample: more specifically, we have removed each day of each week at a time, hence producing $7$ weeks of $6$ days each out of $1$ week of $7$ days; upon doing so, we have moved from $\simeq150$ snapshots to $\simeq1000$ snapshots. Here, we have focused on a balanced panel of $N=73$ banks.

The performance of the model in reproducing the chosen empirical quantities across the weeks constituting our dataset is, instead, shown in fig.~\ref{fig:appc2}. Its right panel deserves to be commented more. Let us notice the peak in correspondence of the year $2008$: again, the crisis reveals itself as an event challenging the model adopted to describe the system under consideration, letting the error associated to the estimates of the degrees rise.

Let us, now, briefly comment on the possible priors. First, let us carry out a consistency check, by noticing that employing the \emph{deterministic prior} $\pi(z)=\delta_{z,z^*_\text{ML}}$ lets the traditional dcGM be recovered. In order to discuss other choices, let us put ourselves in the sparse-case regime, defined by the position $p_{ij}\simeq zs_is_j$: this leads to

\begin{align}
P(\mathbf{A}_{t+1}|\mathbf{A}_t)&\simeq\frac{\int_0^{+\infty}\left[\prod_{i=1}^{N_{t+1}}\prod_{j(>i)}\left(zs_i^{t+1}s_j^{t+1}\right)^{a_{ij}^{t+1}}\right]\left[\prod_{i=1}^{N_t}\prod_{j(>i)}\left(zs_i^ts_j^t\right)^{a_{ij}^t}\right]\pi(z)dz}{\int_0^{+\infty}\left[\prod_{i=1}^{N_t}\prod_{j(>i)}\left(zs_i^ts_j^t\right)^{a_{ij}^t}\right]\pi(z)dz}\nonumber\\
&\simeq\frac{\prod_{i=1}^{N_{t+1}}\prod_{j(>i)}\left(s_i^{t+1}s_j^{t+1}\right)^{a_{ij}^{t+1}}\cdot\prod_{i=1}^{N_t}\prod_{j(>i)}\left(s_i^ts_j^t\right)^{a_{ij}^t}\int_0^{+\infty}z^{L_{t+1}+L_t}\pi(z)dz}{\prod_{i=1}^{N_t}\prod_{j(>i)}\left(s_i^ts_j^t\right)^{a_{ij}^t}\int_0^{+\infty}z^{L_t}\pi(z)dz}
\end{align}
that can be further simplified into

\begin{align}
P(\mathbf{A}_{t+1}|\mathbf{A}_t)&=\prod_{i=1}^{N_{t+1}}\prod_{j(>i)}\left(s_i^{t+1}s_j^{t+1}\right)^{a_{ij}^{t+1}}\frac{\int_0^{+\infty}z^{L_{t+1}+L_t}\pi(z)dz}{\int_0^{+\infty}z^{L_t}\pi(z)dz};
\end{align}
instantiating the formula above with a \emph{uniform prior}, ranging between $0$ and $\hat{z}<+\infty$, leads us to

\begin{align}
P(\mathbf{A}_{t+1}|\mathbf{A}_t)&=\prod_{i=1}^{N_{t+1}}\prod_{j(>i)}\left(s_i^{t+1}s_j^{t+1}\right)^{a_{ij}^{t+1}}\cdot\frac{\hat{z}^{L_{t+1}+L_t+1}}{L_{t+1}+L_t+1}\cdot\frac{L_t+1}{\hat{z}^{L_t+1}}\simeq\prod_{i=1}^{N_{t+1}}\prod_{j(>i)}\left(s_i^{t+1}s_j^{t+1}\right)^{a_{ij}^{t+1}}\cdot\frac{L_t}{L_{t+1}+L_t}\cdot\hat{z}^{L_{t+1}}.
\end{align}

The marginal probability, instead, reads

\begin{align}
q_{ij}^{t+1}&=\int_0^{+\infty}p_{ij}^{t+1}(z)\frac{P(\mathbf{A}_t|z)\pi(z)}{P(\mathbf{A}_t)}\,dz\nonumber\\
&\simeq\int_0^{+\infty}(zs_i^{t+1}s_j^{t+1})\frac{z^{L_t}\pi(z)}{\int_0^{+\infty}z^{L_t}\pi(z)\,dz}\,dz\nonumber\\
&=s_i^{t+1}s_j^{t+1}\frac{\int_0^{+\infty}z^{L_t+1}\pi(z)\,dz}{\int_0^{+\infty}z^{L_t}\pi(z)\,dz}\nonumber\\
&=s_i^{t+1}s_j^{t+1}\frac{\int_0^{\hat{z}}z^{L_t+1}\,dz}{\int_0^{\hat{z}}z^{L_t}\,dz}\nonumber\\
&=s_i^{t+1}s_j^{t+1}\cdot\frac{L_t+1}{L_t+2}\cdot \hat{z},
\end{align}
an expression confirming that, in the sparse-case regime, a uniform prior affects all the estimates of interest through a common multiplicative factor.

For what concerns the link prediction task, the expression above embodies the so-called \emph{preferential attachment} recipe: since the rank of each link is determined by the (score individuated by the) product of the strengths of the corresponding nodes, such a recipe is independent from the choice of the prior. For what concerns the network reconstruction task, instead, this is no longer true and $z$ must be numerically evaluated. For consistency with the remainder of the analysis, we adopt the ML recipe, prescribing to solve the equation $\sum_{i=1}^N\sum_{j(>i)}zs_is_j=L^*$: in symbols, $z_\text{CL}^*=L^*/\sum_{i=1}^N\sum_{j(>i)}s_is_j$. The upper bound $\hat{z}$ can be determined by identifying the largest $z_\text{CL}^*$ value over all snapshots.

\begin{figure*}[t!]
\centering
\includegraphics[width=0.9\linewidth]{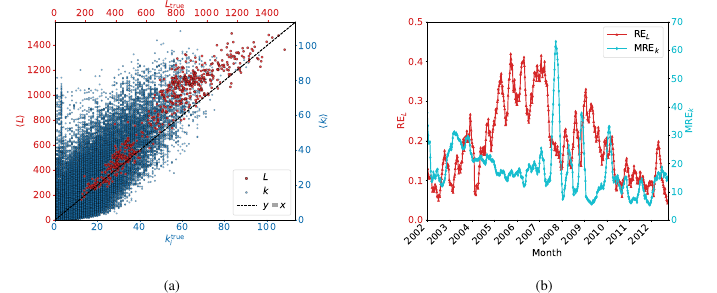}
\caption{\textbf{Performance of the Bayesian Chung-Lu model.} Panel (a): empirical values of the total number of links (red) and node degrees (blue) scattered versus the corresponding predicted ones, pooled across the weeks constituting our dataset; the dashed line marks the identity. A clear upward bias is visible, with most points lying above the diagonal, indicating that the BCLM tends to overestimate both the total number of links and node degrees. Panel (b): evolution of the relative error on the total number of links (red) and the maximum relative error on the node degrees (blue), across the weeks constituting our dataset. Overall, the BCLM yields systematically larger errors than the BFM, thus highlighting the importance of a careful choice of the prior.}
\label{fig:appc3}
\end{figure*}

The performance of what may be called Bayesian Chung-Lu Model (BCLM) is depicted in fig.~\ref{fig:appc3}: as expected, the BFM outperforms it in reproducing the considered quantities. The worse performance of the BCLM is due to both its functional form and the choice of the prior: the first one does not guarantee that all $p_{ij}$s range between $0$ and $1$; the second one, instead, requires the knowledge of the total number of links across all snapshots - an amount of information that is ultimately unnecessary, as proven by our exercise about `self-sustained' inference.

\clearpage

For the sake of completeness, let us notice that the expression of $q_{ij}^{t+1}$ can be rewritten as

\begin{align}
q_{ij}^{t+1}
&=\int_0^{+\infty}p_{ij}^{t+1}(z)\frac{P(\mathbf{A}_t|z)\pi(z)}{P(\mathbf{A}_t)}dz\nonumber\\
&=\int_0^{+\infty}\left(\frac{zs_i^{t+1}s_j^{t+1}}{1+zs_i^{t+1}s_j^{t+1}}\right)\frac{P(\mathbf{A}_t|z)\pi(z)dz}{P(\mathbf{A}_t)}\nonumber\\
&=\int_0^{+\infty}\left(\frac{zs_i^{t+1}s_j^{t+1}}{1+zs_i^{t+1}s_j^{t+1}}\right)\frac{\prod_{m=1}^{N_t}\prod_{n(>m)}\left(\frac{zs_m^ts_n^t}{1+zs_m^ts_n^t}\right)^{a_{mn}^t}\left(\frac{1}{1+zs_m^ts_n^t}\right)^{1-a_{mn}^t}\pi(z)dz}{\int_0^{+\infty}\prod_{m=1}^{N_t}\prod_{n(>m)}\left(\frac{zs_m^ts_n^t}{1+zs_m^ts_n^t}\right)^{a_{mn}^t}\left(\frac{1}{1+zs_m^ts_n^t}\right)^{1-a_{mn}^t}\pi(z)dz}\nonumber\\
&=\int_0^{+\infty}\left(\frac{zs_i^{t+1}s_j^{t+1}}{1+zs_i^{t+1}s_j^{t+1}}\right)\frac{\prod_{m=1}^{N_t}\prod_{n(>m)}\frac{\left(zs_m^ts_n^t\right)^{a_{mn}^t}}{1+zs_m^ts_n^t}\pi(z)dz}
{\int_0^{+\infty}\prod_{m=1}^{N_t}\prod_{n(>m)}\frac{\left(zs_m^ts_n^t\right)^{a_{mn}^t}}{1+zs_m^ts_n^t}\pi(z)dz}\nonumber\\
&=\int_0^{+\infty}\left(\frac{s_i^{t+1}s_j^{t+1}}{1+zs_i^{t+1}s_j^{t+1}}\right)\frac{z^{L_t+1}\left[\prod_{m=1}^{N_t}\prod_{n(>m)}\left(s_m^ts_n^t\right)^{a_{mn}^t}\right]\left[\prod_{m=1}^{N_t}\prod_{n(>m)}\frac{1}{1+zs_m^ts_n^t}\right]\pi(z)dz}{\int_0^{+\infty}z^{L_t}\left[\prod_{m=1}^{N_t}\prod_{n(>m)}\left(s_m^ts_n^t\right)^{a_{mn}^t}\right]\left[\prod_{m=1}^{N_t}\prod_{n(>m)}\frac{1}{1+zs_m^ts_n^t}\right]\pi(z)dz}\nonumber\\
&=\int_0^{+\infty}\left(\frac{s_i^{t+1}s_j^{t+1}}{1+zs_i^{t+1}s_j^{t+1}}\right)\frac{z^{L_t+1}\left[\prod_{m=1}^{N_t}\prod_{n(>m)}\frac{1}{1+zs_m^ts_n^t}\right]\pi(z)dz}{\int_0^{+\infty}z^{L_t}\left[\prod_{m=1}^{N_t}\prod_{n(>m)}\frac{1}{1+zs_m^ts_n^t}\right]\pi(z)dz};
\end{align}
once the strength sequence at time $t$ is given, the dependence on the observed binary snapshot is summarized by the sole scalar quantity $L_t$ - in other words, knowing the entire adjacency matrix is not required.

\clearpage

\counterwithin*{figure}{section}
\stepcounter{section}
\renewcommand{\thefigure}{D.\arabic{figure}}

\section*{APPENDIX D.\\Numerical integration of the posterior predictive distribution}
\hypertarget{AppD}{}

Let us start by recalling that the probability distribution of the fitness model reads

\begin{align}
P(\mathbf{A}_t|z)&=\prod_{i=1}^{N_t}\prod_{j(>i)}\left(p_{ij}^t\right)^{a_{ij}^t}\left(1-p_{ij}^t\right)^{1-a_{ij}^t}=\prod_{i=1}^{N_t}\prod_{j(>i)}\left(\frac{zs_i^ts_j^t}{1+zs_i^ts_j^t}\right)^{a_{ij}^t}\left(\frac{1}{1+zs_i^ts_j^t}\right)^{1-a_{ij}^t}=\prod_{i=1}^{N_t}\prod_{j(>i)}\frac{\left(zs_i^ts_j^t\right)^{a_{ij}^t}}{1+zs_i^ts_j^t};
\end{align}
its logarithm, then, gives the log-likelihood

\begin{align}
\mathcal{L}(z)&=\sum_{i=1}^{N_t}\sum_{j(>i)}a_{ij}^t\ln z+\sum_{i=1}^{N_t}\sum_{j(>i)}a_{ij}^t\ln[s_i^ts_j^t]-\sum_{i=1}^{N_t}\sum_{j(>i)}\log[1+zs_i^ts_j^t]\nonumber\\
&=L_t\:\ln z+\sum_{i=1}^{N_t}\sum_{j(>i)}a_{ij}^t\ln[s_i^ts_j^t]-\sum_{i=1}^{N_t}\sum_{j(>i)}\log[1+zs_i^ts_j^t].
\end{align}

\subsection*{Log-space formulation}

Let us, now, set $u=\ln z$, a transformation mapping the original domain $(0,+\infty)$ onto $\mathbb{R}$ and preventing numerical overflow when $z$ becomes large. In log-space, the expression above becomes

\begin{equation}
\ell(u)=L_t\:u-\sum_{i=1}^{N_t}\sum_{j(>i)}\ln[1+e^us_i^t s_j^t],
\end{equation}
where we have dropped the terms that do not depend on $z$. Now, since $p(u|\mathbf{A}_t)du=p(z|\mathbf{A}_t)dz$, one has

\begin{equation}
p(u|A_t)=p(z|\mathbf{A}_t)\frac{dz}{du}=p(\mathbf{A}_t|z)\pi(z)\frac{dz}{du}=P(\mathbf{A}_t|e^u)\pi(e^u)e^u;
\end{equation}
hence, the log-posterior reads

\begin{equation}
g(u)=\ell(u)+\ln\pi(e^u)+u
\end{equation}
(where we have dropped the terms that do not depend on $u$). The $(i,j)$ term of the predictive probability distribution can be, thus, written as

\begin{equation}\label{eq:upost}
q_{ij}^{t+1}=\int_{-\infty}^{+\infty}\left(\frac{e^us_i^{t+1}s_j^{t+1}}{1+e^us_i^{t+1}s_j^{t+1}}\right)\frac{e^{g(u)}}{\int e^{g(v)}dv}du.
\end{equation}

\subsection*{Gauss-Hermite quadrature}

The Gauss-Hermite quadrature scheme works by posing

\begin{equation}
g(u)\simeq g(\hat{u})-\frac{(u-\hat{u})^2}{2\sigma^2},
\end{equation}
where $\hat{u}$ denotes the mode of $g(u)$ and $\sigma^2=[-g''(\hat{u})]^{-1}$ denotes the inverse curvature at the mode. By introducing the rescaled variable $x=(u-\hat{u})/(\sqrt{2}\sigma)$, the $(i,j)$ term of the predictive probability distribution becomes

\begin{equation}
q_{ij}^{t+1}\simeq\sum_{k=1}^{K}w_k\frac{e^{\hat{u}+\sqrt{2}\sigma x_k}s_i^{t+1}s_j^{t+1}}{1+e^{\hat{u}+\sqrt{2}\sigma x_k}s_i^{t+1}s_j^{t+1}}
\end{equation}
where $\{x_k,w_k\}_{k=1}^{K}$ denote the Gauss-Hermite nodes and weights.

\begin{figure*}[t!]
\centering
\includegraphics[width=0.9\linewidth]{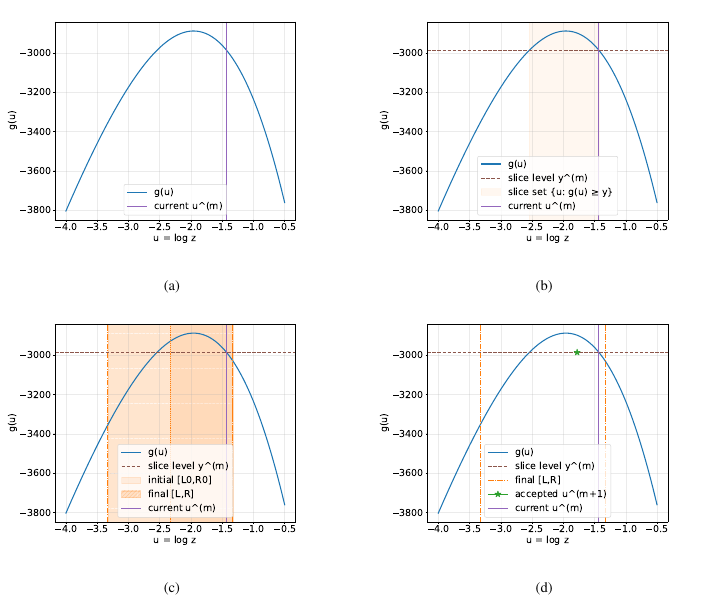}
\caption{\textbf{Illustration of one iteration of the `slice-sampling' algorithm applied to the posterior $g(u)$.} Panel (a): non-normalized log-posterior and current state $u^{(m)}$. Panel (b): the horizontal `slice' $y^{(m)}=g\left(u^{(m)}\right)$ defines the set $\{u:g(u)\geq y^{(m)}\}$. Panel (c): initial bracket $[L_0,R_0]$ of nominal width $w$ and enlarged bracket $[L,R]$ from the stepping-out phase. Panel (d): possible $u^*$ values are drawn from $\text{Unif}(L,R)$ until $g(u^*)\geq y^{(m)}$, thus yielding the next state $u^{(m+1)}$. In this example, the prior on $z=e^u$ is $\text{Gamma}(k,\theta)$, calibrated on the weeks constituting the years $1999$-$2001$, and the posterior is computed for the last week of the year $2001$.}
\label{fig:appd1}
\end{figure*}

\subsection*{Slice sampling}

A complementary integration scheme is represented by slice sampling. Since the normalization constant of the posterior is unknown, direct sampling from $p(u|\mathbf{A}_t)\propto e^{g(u)}$ is not possible: slice sampling, then, introduces an auxiliary variable $y$ representing a random vertical level below the non-normalized posterior curve $e^{g(u)}$. At the $m$-th iteration, we draw $r^{(m)}\sim\text{Unif}(0,1)$, so that

\begin{equation}
e^{y^{(m)}}=e^{g\left(u^{(m)}\right)}r^{(m)}
\end{equation}
defines a horizontal `slice' at a random height uniformly distributed between $0$ and the current density value $e^{g\left(u^{(m)}\right)}$: the point $u^{(m+1)}$ is, then, drawn uniformly from those whose posterior density lies above this `slice', i.e. from the set

\begin{equation}
S^{(m)}=\{u:g(u)\geq y^{(m)}\};
\end{equation}
the accepted draw lies uniformly within the `slice'. Repeating the two-step procedure moves the pair $(u^{(m)},y^{(m)})$ uniformly within the area under the posterior, so that the marginal samples $\{u^{(m)}\}$ follow the desired distribution. After a burn-in phase, the draws are mapped back to $z^{(m)}=e^{u^{(m)}}$. Since

\begin{figure}[t!]
\centering
\includegraphics[width=0.9\linewidth]{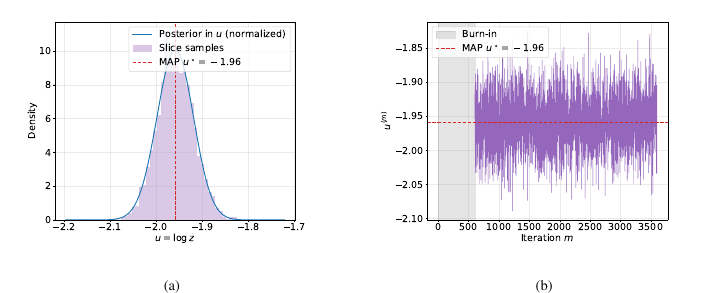}
\caption{\textbf{Posterior diagnostics for slice sampling.} Panel (a): numerically normalized posterior distribution corresponding to the week $\#52$ of the year $2001$, superimposed to the histogram of `slice-sampled' draws of $u$; the vertical line marks the MAP estimate. Panel (b): evolution of the `slice-sampling' chain across iterations (the number of sampled values amounts to $M=3000$, after a burn-in phase of $600$ steps), rapidly stabilizing around the higher-density region of the posterior distribution.}
\label{fig:appd2}
\end{figure}

\begin{equation}
p_{ij}^{t+1}=\frac{e^us_i^{t+1}s_j^{t+1}}{1+e^us_i^{t+1}s_j^{t+1}}
\end{equation}
and the BFM averages it over the posterior distribution $\propto e^{g(u)}$, one gets the same expression as in eq.~\ref{eq:upost}. Knowing the denominator $\int e^{g(v)}dv$, representing the (unknown) normalization constant of the posterior, is, however, not necessary, as $q_{ij}^{t+1}$ is estimated as

\begin{equation}
q_{ij}^{t+1}\simeq\frac{1}{M}\sum_{m=1}^M\frac{e^{u^{(m)}}s_i^{t+1}s_j^{t+1}}{1+e^{u^{(m)}}s_i^{t+1}s_j^{t+1}}.
\end{equation}

Figure~\ref{fig:appd1} illustrates the mechanics of one iteration of the so-called `slice-sampling' algorithm applied to the posterior $g(u)$: `slice-sampling' scales effectively, as it does not require computing the normalization constant. Figure~\ref{fig:appd2} complements our illustration, showing both the shape of the numerically normalized posterior distribution corresponding to the week $\#52$ of the year $2001$ and the evolution of the `slice-sampling' chain across iterations.

\clearpage

\section*{APPENDIX E.\\DATA DESCRIPTION}
\hypertarget{AppE}{}

We employ transaction-level records from the Electronic Market for Interbank Deposits (eMID), a screen-based market for unsecured deposits~\cite{marzi2026reproducing,macchiati2025spectral,IoriOvernightMoneyMarket2008,FingerFrickeLux2012}. We restrict the analysis to overnight transactions, which account for the vast majority of activity on the platform. Our sample spans trading days from January $1999$ to September $2012$. Each trading day $d$ defines a weighted, directed matrix $\mathbf{V}(d)$ whose entry $v_{ij}(d)$ equals the total notional amount lent by bank $i$ to bank $j$ on day $d$, with $v_{ii}(d)=0$. Trading occurs exclusively on business days, so weekends and bank holidays are absent from the raw records.
Throughout the present work, data are aggregated at the weekly level. Weeks correspond to ISO calendar weeks (Monday to Sunday).

Let $\Delta_t$ denote the set of trading days belonging to week $t$. Since the number of business days varies across weeks, aggregation is performed over the set of trading days effectively observed within each window. Weekly weights are computed as

\begin{equation}
v_{ij}^{(t)}=\sum_{d\in\Delta_t}v_{ij}(d).
\end{equation}

For each week $t$, we restrict the node set to banks that are active during $\Delta_t$, i.e. involved in at least one transaction in that week. The number of active banks is, therefore, time-dependent. For example, during the period $1999$-$2002$, the weekly number of active institutions ranges between $136$ and $197$, with an average close to $163$ banks per week.

Since empirical records are weighted and directed, we construct an undirected exposure matrix by symmetrizing weekly weights as

\begin{equation}
w_{ij}^{(t)}=v_{ij}^{(t)}+v_{ji}^{(t)},\quad i\neq j,
\end{equation}
and define the corresponding binary adjacency matrix

\begin{equation}
a_{ij}^{(t)}=\mathbbm{1}\{w_{ij}^{(t)}>0\},\quad a_{ii}^{(t)}=0,
\end{equation}
so that, in matrix notation, $\mathbf{A}_t=\Theta[\mathbf{W}_t>0]$ element-wise. Node strengths are computed from symmetrized weights as

\begin{equation}
s_i^{(t)}=\sum_{j(\neq i)}w_{ij}^{(t)}
\end{equation}
and are used as exogenous fitnesses informing the dcGM and related models.

\clearpage

\counterwithin*{figure}{section}
\stepcounter{section}
\renewcommand{\thefigure}{F.\arabic{figure}}

\section*{APPENDIX F.\\Marginal probability induced by the estimated prior}
\hypertarget{AppF}{}

\begin{figure*}[b!]
\centering
\includegraphics[width=0.85\linewidth]{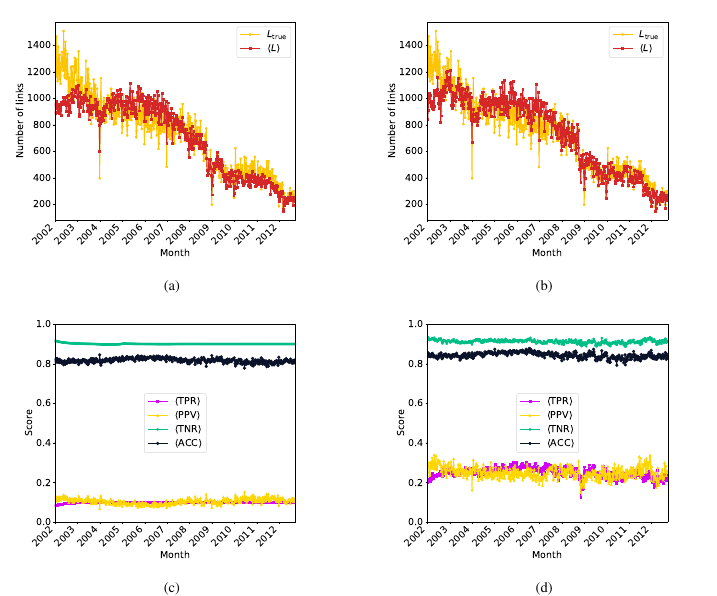}
\caption{\textbf{Self-sustained inference under the BERM and the BFM.} Panels (a) and (b): evolution of the empirical (yellow) and expected (red) number of links under the self-sustained BERM and BFM, respectively. Panels (c) and (d): corresponding evolution of $\langle\text{TPR}\rangle$, $\langle\text{PPV}\rangle$, $\langle\text{TNR}\rangle$ and $\langle\text{ACC}\rangle$. Both self-sustained variants reproduce the number of links and achieve a large $\langle\text{ACC}\rangle$ score, driven by the large value of $\langle\text{TNR}\rangle$; the BFM, however, outperforms the BERM in achieving larger $\langle\text{TPR}\rangle$ and $\langle\text{PPV}\rangle$ scores.}
\label{fig:appf1}
\end{figure*}

The key formula to carry out what we called `self-sustained' inference reads

\begin{align}
r_{ij}^{t+1}&=\int_0^{+\infty}\left(\frac{zs_i^{t+1}s_j^{t+1}}{1+zs_i^{t+1}s_j^{t+1}}\right)\frac{P(\mathbf{Q}_t|z)\pi(z)}{P(\mathbf{Q}_t)}dz=\frac{\int_0^{+\infty}\left(\frac{zs_i^{t+1}s_j^{t+1}}{1+zs_i^{t+1}s_j^{t+1}}\right)\left[\prod_{i=1}^{N_t}\prod_{j(>i)}\frac{\left(zs_i^ts_j^t\right)^{q_{ij}^t}}{1+zs_i^ts_j^t}\right]\pi(z)dz}{\int_0^{+\infty}\left[\prod_{i=1}^{N_t}\prod_{j(>i)}\frac{\left(zs_i^ts_j^t\right)^{q_{ij}^t}}{1+zs_i^ts_j^t}\right]\pi(z)dz},
\end{align}
making it explicit that $\mathbf{Q}_t$ is employed as a prior to infer $\mathbf{R}_{t+1}$. Figure~\ref{fig:appf1} depicts the reliability of such a procedure in reproducing the $\langle\text{TPR}\rangle$, the $\langle\text{PPV}\rangle$, the $\langle\text{TNR}\rangle$ and the $\langle\text{ACC}\rangle$ across the weeks constituting our dataset. Both `self-sustained' variants reproduce the evolution of the number of links over more than a decade of weekly eMID snapshots (2002-2012), despite receiving no topological information beyond that of the initial calibration period ($1999$-$2001$); within such a fully self-sustained regime, the BFM achieves a substantially higher accuracy in reproducing the degree sequence as well, whereas the BERM cannot account for degree heterogeneity.

\end{document}